\documentclass[reprint,superscriptaddress,aps,longbibliography]{revtex4-1}
\usepackage[utf8]{inputenc}
\usepackage[T1]{fontenc}
\usepackage{amsmath}
\usepackage{amssymb}
\usepackage{graphicx}
\usepackage[capitalize, nameinlink]{cleveref}
\usepackage{subfigure}
\graphicspath{{figures/}}
\usepackage{adjustbox}
\usepackage{booktabs}
\usepackage{multirow}
\newcommand{\ra}[1]{\renewcommand{\arraystretch}{#1}}
\usepackage{color}
\usepackage{rotating}
\usepackage{ulem}

\begin{document}

\title{{Symmetry-dependent ultrafast manipulation of nanoscale magnetic domains}}

\author{Nanna Zhou Hagstr\"{o}m}
\affiliation{Department of Physics, Stockholm University, 106 91 Stockholm, Sweden}

\author{Rahul Jangid}
\author{Meera}
\affiliation{Department of Materials Science and Engineering, University of California Davis, CA, USA}
\author{Diego Turenne}
\affiliation{Department of Physics and Astronomy, Uppsala University, Uppsala, Sweden} 
\author{Jeffrey Brock}
\affiliation{Center for Memory and Recording Research, University of California San Diego, La Jolla, CA, USA}
\author{{Erik S. Lamb}}
\author{Boyan Stoychev}
\affiliation{Department of Physics, University of California San Diego, La Jolla, CA, USA}

\author{Justine Schlappa}
\author{Natalia Gerasimova}
\author{Benjamin Van Kuiken}
\author{Rafael Gort}
\author{Laurent Mercadier}
\author{Loïc Le Guyader}
\author{Andrey Samartsev}
\author{Andreas Scherz}
\author{Giuseppe Mercurio}
\affiliation{European XFEL GmbH, Holzkoppel 4, 22869 Schenefeld, Germany}

\author{Hermann {A.} D\"{u}rr}
\affiliation{Department of Physics and Astronomy, Uppsala University, Uppsala, Sweden}
\author{Alexander H. Reid}
\affiliation{{Linac} Coherent Light Source, SLAC National Accelerator Laboratory, {Menlo Park}, CA, USA}
\author{{Monika Arora}}
\affiliation{Quantum Electromagnetics Division, National Institute of Standards and Technology, Boulder, CO, USA}
\author{Hans {T.} Nembach} 
\affiliation{{Department of Physics}, University of Colorado, Boulder, CO, USA}
\affiliation{Quantum Electromagnetics Division, National Institute of Standards and Technology, Boulder, CO, USA}
\author{Justin M. Shaw}  
\affiliation{Quantum Electromagnetics Division, National Institute of Standards and Technology, Boulder, CO, USA}
\author{Emmanuelle Jal}
\affiliation{{Sorbonne Universit\'{e}, CNRS, Laboratoire de Chimie Physique – Mati\`{e}re et Rayonnement, LCPMR, Paris 75005, France}}
\author{Eric E. Fullerton}
\affiliation{Center for Memory and Recording Research, University of California San Diego, La Jolla, CA, USA}
\author{Mark W. Keller}
\affiliation{Quantum Electromagnetics Division, National Institute of Standards and Technology, Boulder, CO, USA}
\author{Roopali Kukreja}
\affiliation{Department of Materials Science and Engineering, University of California Davis, CA, USA}
\author{Stefano Bonetti}
\affiliation{Department of Physics, Stockholm University, 106 91 Stockholm, Sweden}
\affiliation{Department of Molecular Sciences and Nanosystems, Ca' Foscari University of Venice, 30172 Venezia, Italy}
\author{Thomas J. Silva}
\affiliation{Quantum Electromagnetics Division, National Institute of Standards and Technology, Boulder, CO, USA}
\author{Ezio Iacocca}
\affiliation{Department of Mathematics, Physics, and Electrical Engineering, Northumbria University, Newcastle upon Tyne, NE1 8ST, United Kingdom}
\affiliation{Center for Magnetism and Magnetic Materials, University of Colorado Colorado Springs, Colorado Springs, CO, USA}

\date{\today}

\begin{abstract}
Symmetry is a powerful concept in physics, but its applicability to far-from-equilibrium states is still being understood. Recent attention has focused on how far-from-equilibrium states lead to spontaneous symmetry breaking. Conversely, ultrafast optical pumping can be used to drastically change the energy landscape and quench the magnetic order parameter in magnetic systems. Here, we find a distinct {symmetry-dependent} ultrafast behaviour by use of ultrafast {x-ray scattering from} magnetic patterns with varying degrees of isotropic and anisotropic symmetry. {After pumping with an optical laser, the scattered intensity reveals} a radial shift exclusive to the isotropic component and {exhibits} a faster recovery time from quenching for the anisotropic component. These features {arise} even when both symmetry components are concurrently measured, {suggesting} a correspondence between the excitation and the magnetic order symmetry. Our results underline the importance of symmetry as a critical variable to manipulate the magnetic order in the ultrafast regime.
\end{abstract}
\maketitle

Ultrafast manipulation of symmetry is achievable in a wide variety of physical systems that rely on non-equilibrium pathways to access hidden states in their energy landscape. Far-from-equilibrium transitions from symmetric to symmetry-broken states have been observed, e.g. by photo-induced superconductivity~\cite{budden2021evidence}, structural modification of alloys~\cite{hase2015femtosecond}, manipulation of topological phases in Weyl semimetals~\cite{sie2019ultrafast}, vibrational dynamics following melting of atomic charge order in nickelates~\cite{coslovich2017ultrafast}, hidden states during spontaneous symmetry breaking of charge density waves~\cite{zhou2021nonequilibrium}, and charge separation of chiral organic molecules~\cite{wu2015ultrafast}. Symmetry can be also manipulated in magnetic materials because of the interplay between their local and nonlocal order parameters. Recent studies have indeed demonstrated that the homogeneously magnetized ferrimagnet GdFeCo undergoes phase-ordering kinetics through the ultrafast formation of localized defects~\cite{iacocca2019spincurrentmediated}. Moreover, topological phases could be accessed in ferromagnetic materials biased with an external magnetic field, demonstrating the picosecond emergence and subsequence stabilization of skyrmion lattices~\cite{buttner2021observation} and in ferrimagnets showing the transition from helical to skyrmion phases~\cite{berruto2018laserinduced}. In these works, the manipulation of symmetry occurred within the magnetic or spin degree of freedom. However, ultrafast excitation of metallic magnetic materials~\cite{kirilyuk2010ultrafasta} also induces non-equilibrium spin currents~\cite{graves2013nanoscale} that can produce torques~\cite{balaz2020domain} and affect the picosecond dynamics of the spin degree of freedom~\cite{beaurepaire1996ultrafast, koopmans2010explaining, turgut2013controlling,  pfau2012ultrafast, vodungbo2012laserinduced, tengdin2018critical, hennes2020laserinduced, zusin2020ultrafast}.

\begin{figure*}[ht]
\begin{center}
\includegraphics[width=\textwidth]{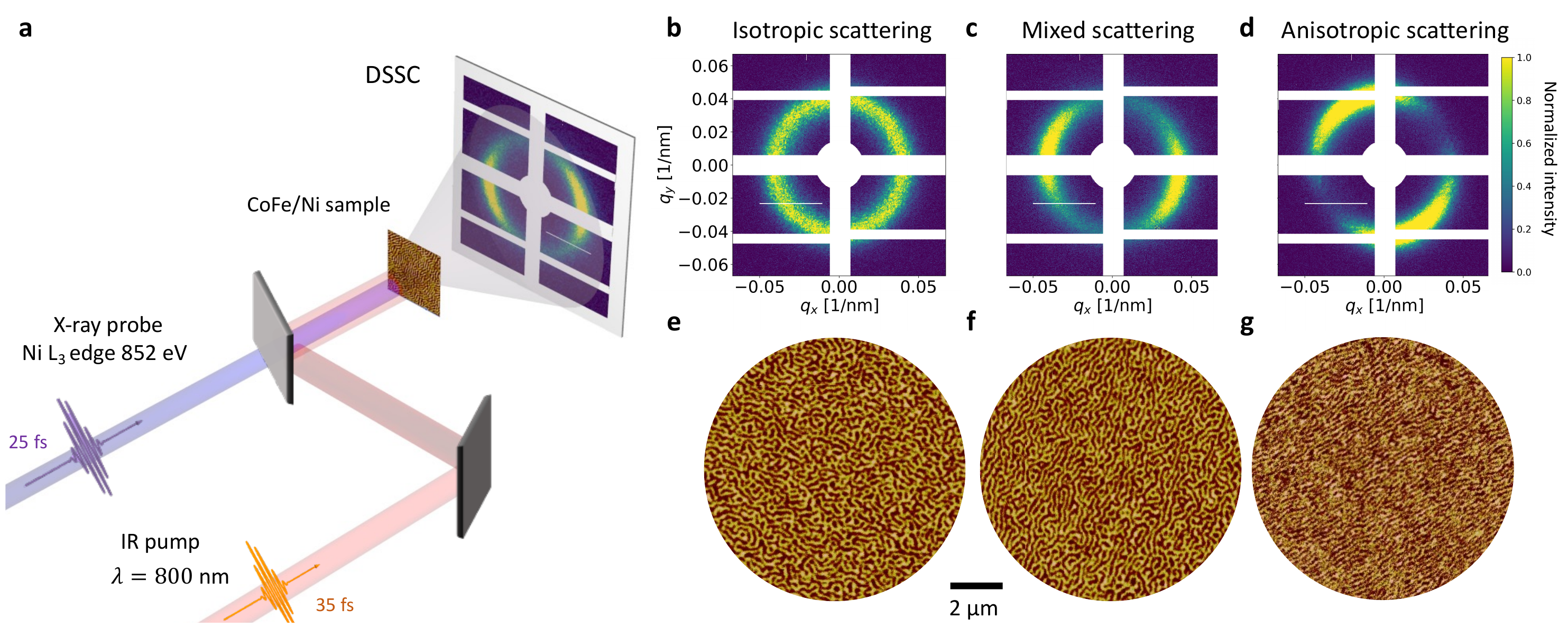}
\caption{\textbf{Scattering from magnetic domain patterns.}  \textbf{a} Schematic of the experimental pump-probe setup. The sample is excited by IR pulses and probe by linearly polarized X-rays tuned to the L$_3$-edge of Ni. The scattered photons are collected on the DSSC detector. Representative diffraction patterns are shown for \textbf{b} labyrinth, \textbf{c} mixed, and \textbf{d} stripe domain patterns. Corresponding $10\times10$~$\mu$m$^2$ MFM real-space images are shown in \textbf{d}-\textbf{f}, illustrating the varying degree of randomness for each domain pattern.}
\label{fig:setup_SAXS_MFM}
\end{center}
\end{figure*}

A clear manifestation of spin-current-induced ultrafast magnetization dynamics is found in materials exhibiting a domain pattern. It was recognized that materials stabilized in a stripe domain pattern could be demagnetized more quickly than the uniformly magnetized sample~\cite{vodungbo2012laserinduced}. This work provided a first indication of symmetry-dependent ultrafast quench of the local magnetic order parameter. It is then natural to inquire how the different symmetries in domain patterns affect the picosecond magnetization dynamics and how can the symmetries themselves be manipulated by optical excitations. Such control of magnetism at the femtosecond timescales is particularly important by the demand for energy efficient and fast magnetic storage devices \cite{zutic2004spintronics}.

To study the far-from-equilibrium dynamics of these spatial symmetries occurring at nanoscale, time-resolved X-ray scattering from free electron lasers remains the preferred method to achieve the necessary combined temporal and spatial resolution~\cite{pfau2012ultrafast, vodungbo2012laserinduced, graves2013nanoscale, iacocca2019spincurrentmediated, hennes2020laserinduced, zusin2020ultrafast, jeppson2021capturing}. The detected scattered intensity pattern directly correlates to the domain pattern symmetry. {Stripe domains exhibit a preferential spatial translational symmetry in one dimension, leading to a distinctive anisotropic scattering pattern. Conversely, labyrinth domains exhibit no spatial translational symmetry, leading to an isotropic scattering ring.} Studies in both stripe and labyrinth domain patterns have provided a wealth of observations that to date remain disparate and controversial. Initial studies on labyrinth domains in Co/Pt multilayers reported a ring contraction that was interpreted as a result of spin-current induced domain-wall broadening~\cite{pfau2012ultrafast}. Later studies in CoFe/Ni multilayers that could access higher order diffraction rings were able to disentangle domain-wall broadening from the periodicity, suggesting an ultrafast domain rearrangement~\cite{zusin2020ultrafast}. More recently, a similar shift in the scattered ring was observed for chiral labyrinth domains~\cite{kerber2020faster,leveille2021ultrafast}. This shift, however, has not been observed in stripe domains~\cite{vodungbo2012laserinduced,hennes2020laserinduced,fan2019timeresolving}, apparently contradicting earlier works.

We resolve these controversies by studying the time-resolved X-ray scattering from magnetic multilayers exhibiting mixed stripe and labyrinth domain characteristics. By isolating the scattering produced by each symmetry component, we demonstrate symmetry-dependent ultrafast dynamics of the long-range magnetic order. In particular, only the isotropic symmetry component exhibits a {peak} shift in its scattered {pattern}, even when both symmetry components are present at the same time. This eliminates the possibility of artifacts in the measurements and its relationship to domain-wall broadening. We also detect a linear dependence in the recovery time constant to the quench, with a symmetry-dependent characteristic speed. This result points to a further dependency between symmetry and disorder as well as insights into the possible mechanisms driving the recovery of the long-range magnetization.

{Time-resolved small angle X-ray scattering experiments (SAXS) are performed on CoFe/Ni multilayers grown on Si membranes (see methods) at the European X-ray free electron laser (EuXFEL)}. \Cref{fig:setup_SAXS_MFM}\textbf{a} shows the {pump-probe }schematic of the experimental setup. The pump laser is synchronized with the FEL at half of the X-ray probe pulses frequency to collect both the scattering data from ultrafast dynamics (``pumped'') and in quasi-equilibrium (``unpumped'') within the same measurement run. More details on the experimental setup are provided in the methods. We note that the white regions in the scattering correspond to non-active or faulty areas of the DSSC detector.

Representative examples of {static} scattering patterns are shown in \Cref{fig:setup_SAXS_MFM}\textbf{b}-\textbf{d}. This data is collected within the same membrane but with the X-ray beam illuminating different areas of it. Each of these patterns illustrate distinct long-range symmetries present in the membrane which were then characterized in real space using magnetic force microscopy (MFM) as shown in panels \textbf{e}-\textbf{g}. The isotropic ring (\textbf{b}) results from labyrinth domains (\textbf{e}) {while the} anisotropic, lobed pattern (\textbf{d}) results from domains {with a similar translational symmetry to stripe domains} (\textbf{g}). We note that we do not use an external in-plane field to induce a stripe domain pattern~\cite{hellwig2003xray,vodungbo2012laserinduced,hennes2020laserinduced}, but instead find this preferential orientation in sample areas that are likely subject to strain (see Supplementary Note~\ref{sec:SIstrain}). {We also observe a ``mixed'' state where both symmetry features are visible in the scattering } (\textbf{c}). The corresponding MFM image (\textbf{f}) indicates that this pattern arises from a varying degree randomness in the spatial periodicity of the domain structure. 
\begin{figure}[t]
\begin{center}
\includegraphics[width=\columnwidth]{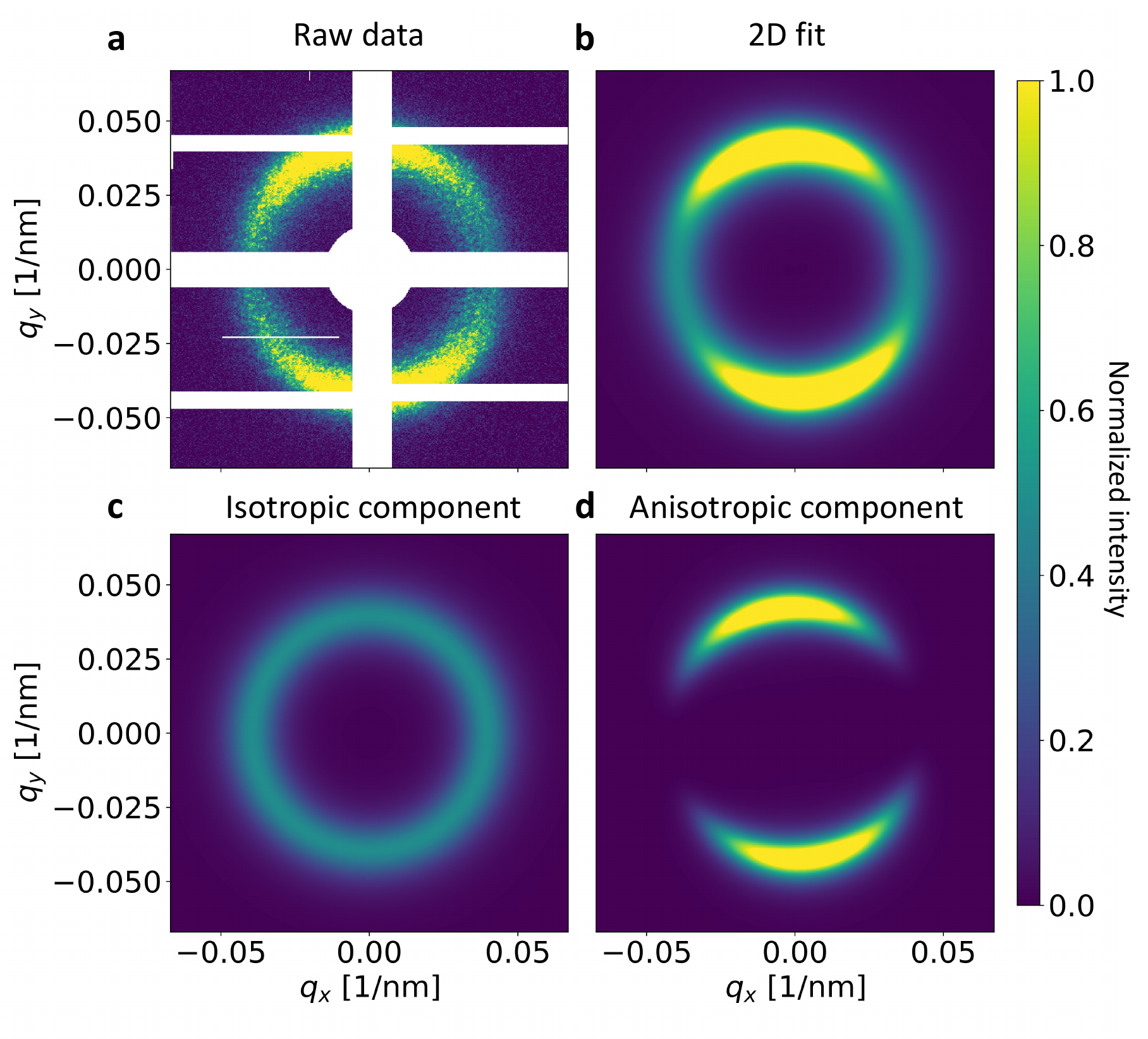}
\caption{\textbf{Two dimensional fit of the scattering data.} \textbf{a} The raw experimental data suffers from missing intensity in the non-active regions of the DSSC. \textbf{b} Two dimensional fit with \cref{eq:2Dfit}. The fit allows to fully separate the \textbf{c} isotropic and the \textbf{d} anisotropic components of the scattering.}
\label{fig:2Dfit}
\end{center}
\end{figure}

\begin{figure}[t]
\begin{center}
\includegraphics[width=\columnwidth]{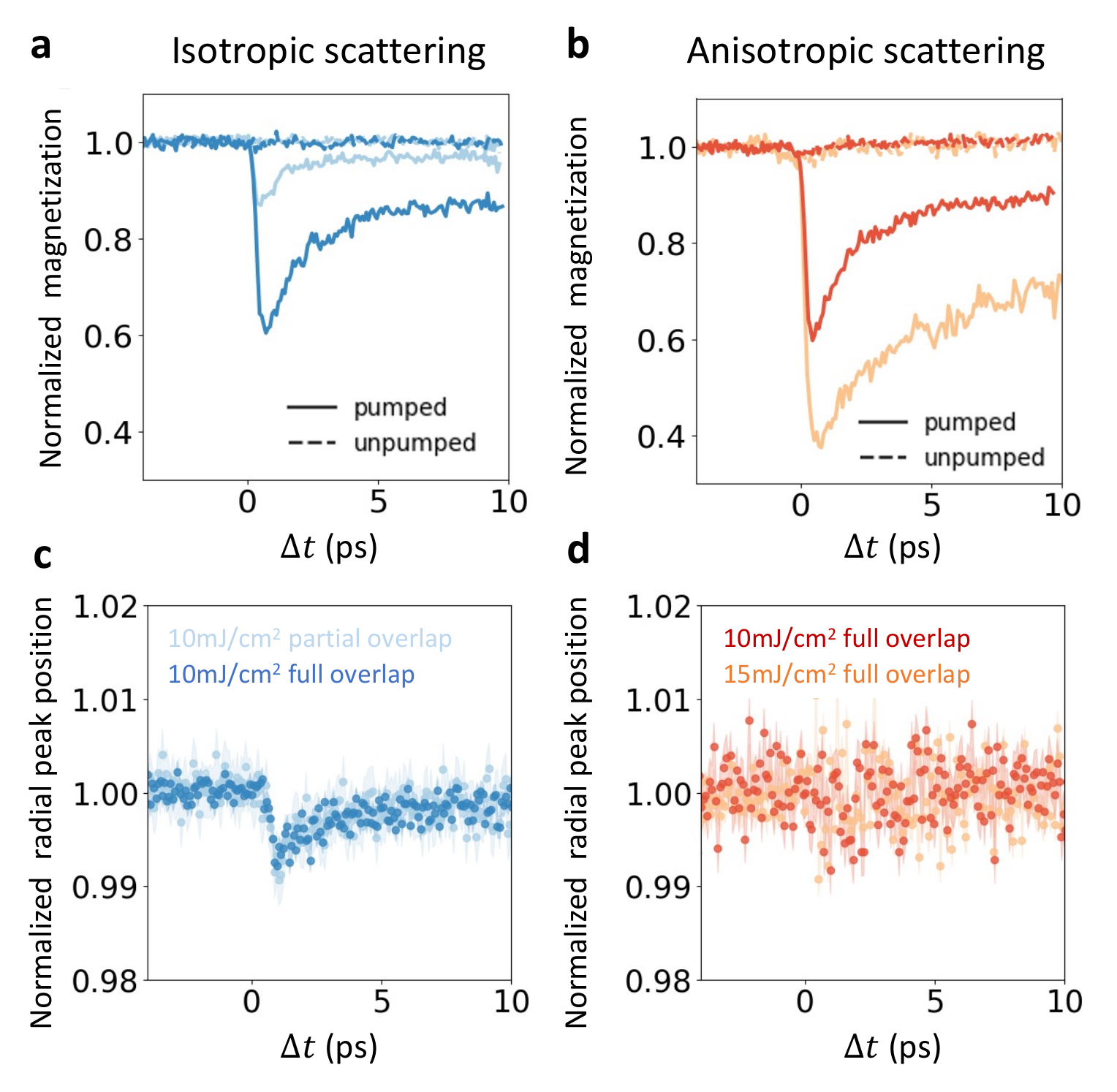}
\caption{\textbf{Ultrafast dynamics of isotropic and anisotropic scattering patterns.} Two-dimensional fitting of the full diffraction pattern is used to extract the {time traces of} both the demagnetization and the peak position for the isotropic and anisotropic scattering patterns.  The normalized demagnetization of isotropic and anisotropic scattering patterns is shown in \textbf{a} and \textbf{b}, respectively. At a fluence of 10~mJ cm$^{-2}$, the demagnetization is shown by the solid curves and corresponding unpumped data  by dashed curves. In both cases, the demagnetization is similar. For the isotropic {scattering}, the effect of a spatial shift between the pump and probe spots {exhibits} a weaker demagnetization {due to a weaker optical pumping, shown by the semitransparent curves named partial overlap}. For the anisotropic ring, a higher fluence of 15~mJ cm$^{-2}$ results in a higher quench, shown by the semitransparent curves. The corresponding temporal evolution of the scattering peak position is shown in \textbf{c} and \textbf{d} for the isotropic and anisotropic scattering patterns, respectively. Only the isotropic component exhibits a shift in the peak position of approximately 1\% and within error bars{, independently of the pump fluence}. The shaded area corresponds to the standard deviation of the fitted quantities. }
\label{fig:pure_states}
\end{center}
\end{figure}
Traditionally, the time-dependent 2D scattering has been analyzed within a 1D representation achieved by azimuthal integration, either over the isotropic ring~\cite{pfau2012ultrafast, zusin2020ultrafast,leveille2021ultrafast, kerber2020faster} or the anisotropic lobes~\cite{vodungbo2012laserinduced, hennes2020laserinduced}. However, our scattering data precludes this simple analysis, both because of the mixed isotropic-anisotropic scattering and because of the significant amount of data lost to the non-active regions of the detector. To fully analyze the data, we developed a 2D fitting procedure that accurately models the varying degree of domain symmetry in our samples and allows us to reconstruct the full scattering pattern. Motivated by the MFM images, we utilize a fitting function for the scattered intensity given by,
\begin{equation}
    f(\mathbf{q}, \varphi) = B + f_\mathrm{iso}(\mathbf{q}) + f_\mathrm{aniso}(\mathbf{q}, \varphi) \label{eq:2Dfit}
\end{equation}
where $\mathbf{q}$ is the wavevector, $B$ is a uniform background, $f_\mathrm{iso}(\textbf{q})$ is the isotropic component that is a function of the wavevector, and $f_\mathrm{aniso}(\mathbf{q},\varphi)$ is the anisotropic component which is a function of both the wavevector and the azimuthal angle $\varphi$. The functional form for both the isotropic and anisotropic components are given in the methods. 
\begin{table*}[t]
\ra{1.1}
\begin{adjustbox}{width=\textwidth}
\begin{tabular}{lccccccccccc} \toprule
 \multicolumn{2}{c}{} & \multicolumn{5}{c}{Isotropic component} & \multicolumn{5}{c}{Anisotropic component} \\ \cmidrule(lr){3-7}\cmidrule(lr){8-12}
  \multicolumn{2}{c}{}  & \multicolumn{1}{c}{Pre-pumped} & \multicolumn{4}{c}{Pumped} & \multicolumn{1}{c}{Pre-pumped} & \multicolumn{4}{c}{Pumped} \\ \cmidrule(lr){3-3}\cmidrule(lr){4-7}\cmidrule(lr){8-8}\cmidrule(lr){9-12}
Fluence & SAXS pattern & size (nm) & $\Delta q_0/ q_0$ (\%) & $\Delta M / M$ (\%) & t$_\mathrm{min}$ (ps) & $\tau_{R} $ (ps) & size (nm) & $\Delta q_1/ q_1$ (\%) & $\Delta M / M$ (\%) & t$_\mathrm{min}$ (ps) & $\tau_{R}$ (ps) \\ \midrule 
$20$~mJ cm$^{-2}$ FO (day 3) & mixed       & 78.30 $\pm$ 0.11 & 1.0 $\pm$ 0.4          & 43.5 $\pm$ 0.9 & 0.714 $\pm$ 0.033  & 3.16 $\pm$ 0.27 & 74.55 $\pm$ 0.04 & 0                      & 35.4 $\pm$ 1.0 & 0.549 $\pm$ 0.043  & 2.25 $\pm$ 0.14  \\ 
$15$~mJ cm$^{-2}$ FO (day 3) & mixed       & 77.27 $\pm$ 0.14 & 1.5 $\pm$ 0.6          & 29.3 $\pm$ 1.0 & 0.660 $\pm$ 0.039  & 2.80 $\pm$ 0.23 & 73.58 $\pm$ 0.04 & 0                      & 20.9 $\pm$ 1.4 & 0.549 $\pm$ 0.043  & 1.52 $\pm$ 0.14  \\ 
$25$~mJ cm$^{-2}$ FO (day 4) & mixed       & 71.96 $\pm$ 0.34 & 3.5 $\pm$ 2.3          & 38.2 $\pm$ 1.4 & 0.970 $\pm$ 0.049  & 2.05 $\pm$ 0.17 & 74.06 $\pm$ 0.14 & 0.5 $\pm$ 0.5          & 34.9 $\pm$ 1.8 & 0.836 $\pm$ 0.049  & 1.68 $\pm$ 0.15  \\ 
$15$~mJ cm$^{-2}$ FO (day 4) & mixed       & 71.86 $\pm$ 0.28 & 2.7 $\pm$ 2.7          & 21.2 $\pm$ 1.0 & 0.738 $\pm$ 0.012  & 1.73 $\pm$ 0.18 & 74.98 $\pm$ 0.14 & 0                      & 18.0 $\pm$ 1.8 & 0.679 $\pm$ 0.021  & 0.99 $\pm$ 0.15  \\ 
$15$~mJ cm$^{-2}$ FO (day 5) & anisotropic & -                & -                      & -              & -                  & -               & 73.08 $\pm$ 0.01 & 0                      & 56.7 $\pm$ 0.7 & 0.58 $\pm$ 0.02    & 2.86 $\pm$ 0.08  \\ 
$15$~mJ cm$^{-2}$ PO (day 5) & mixed       & 77.56 $\pm$ 0.04 & 1.88 $\pm$ 0.07        & 21.9 $\pm$ 0.7 & 0.610 $\pm$ 0.02   & 1.29 $\pm$ 0.07 & 74.16 $\pm$ 0.03 & 0.23 $\pm$ 0.05        & 24.1 $\pm$ 0.2 & 0.61 $\pm$ 0.02    & 1.04 $\pm$ 0.05  \\ 
$10$~mJ cm$^{-2}$ PO (day 5) & isotropic   & 89.38 $\pm$ 0.02 & 0.70 $\pm$ 0.05        & 12.6 $\pm$ 0.5 & 0.840 $\pm$ 0.02   & 0.87 $\pm$ 0.05 & -                & -                      & -              & -                  & -  \\ 
$10$~mJ cm$^{-2}$ FO (day 5) & isotropic   & 76.66 $\pm$ 0.01 & 0.84 $\pm$ 0.06        & 38.5 $\pm$ 0.8 & 0.964 $\pm$ 0.02   & 1.57 $\pm$ 0.03 & -                & -                      & -              & -                  & -  \\
$10$~mJ cm$^{-2}$ FO (day 5) & anisotropic & -                & -                      & -              & -                  & -               & 72.3 $\pm$ 0.01  & 0                      & 37.6 $\pm$ 0.4 & 0.813 $\pm$ 0.013  & 1.65 $\pm$ 0.02     \\ \bottomrule[1pt]
\end{tabular}
\end{adjustbox}
\caption{\textbf{Extracted parameters from the 2D fit.} Measurements are listed by pump fluence and scattering pattern. We distinguish between full overlap (FO) and partial overlap (PO) between probe and probe.  We report the extracted domain size $\pi/q_i$  (nm) from the pre-pumped signal, maximum shift in radial peak position (\%), maximum demagnetization (\%), and demagnetization recovery time (ps) for all measurements. }
\label{table:quench}
\end{table*}

An example of a 2D scattering reconstruction using Eq.~\eqref{eq:2Dfit} is shown in \cref{fig:2Dfit}. The raw experimental {scattering in panel \textbf{a} exhibits features} obstructed by the inactive areas of the detector. The 2D fit reconstruction is shown in panel \textbf{b}, clearly reproducing the main features of the scattering {and allowing us} to separate the isotropic and anisotropic components, shown in panels \textbf{c} and \textbf{d}, respectively. A detailed quantification of the fit quality {is} described in the Supplementary Note~\ref{SI:fitbenchmark}.
  
This fitting procedure is vital to accurately extract the symmetry-dependent far-from-equilibrium magnetization dynamics. First, it resolves artifacts related to azimuthal averaging of the data due to missing pixels, e.g. the apparent development of a bimodal distribution demonstrated in the Supplementary Note~\ref{SI:fitbenchmark}. {Second, it allows us to estimate the number of photons contributing to each symmetry component in the scattering. Using the extracted photon count, we categorize the scattering patterns as isotropic, anisotropic, or mixed, as detailed in the Methods and in Supplementary Note~\ref{SI:photoncount}.}

Based on the {photon-count} classification, we analyze the ultrafast evolution of the average magnetization{, proportional to the fitted amplitude, and the peak position for isotropic and anisotropic scattering patterns. The temporal evolution of both quantities are fitted by the double-exponential function described in the methods and the Supplementary Note~\ref{SI:demag}. The quantities extracted from the fits are summarized in Table~\ref{table:quench}.}

The magnetization normalized to its average value at $t<0$ is shown in \cref{fig:pure_states}\textbf{a} for isotropic scattering patterns and \textbf{b} for anisotropic scattering patterns. The solid dark curves are {obtained} at the same nominal fluence but different sections of the sample. We observe a similar quench for both types of scattering. The dashed curves correspond to the unpumped data and serves to confirm the negligible evolution of the magnetization in a quasi-equilibrium state. It is worth pointing out that the samples do not return to thermal equilibrium at the repetition rate of the experiment, as further discussed in the Supplementary Note~\ref{SI:heating}.

The effect of pump fluence was investigated in two different ways. For the isotropic scattering, we were able to probe the sample at a 50~$\mu$m offset from the pump spot, which we denote ``partial overlap'' (light blue curve in {\cref{fig:pure_states}\textbf{a})}. We observe a three times weaker quench, consistent with an approximately two times weaker pump fluence given its Gaussian profile. For the anisotropic scattering, we observed a larger quench of the demagnetization with increase in pump fluence. These results are in agreement with previous works \cite{koopmans2010explaining, pfau2012ultrafast, vodungbo2012laserinduced, jal2019singleshot, cardin2020wavelength, liu2021sub15fs} and validates the 2D fitting approach.

\begin{figure}[t]
\begin{center}
\includegraphics[width=\columnwidth]{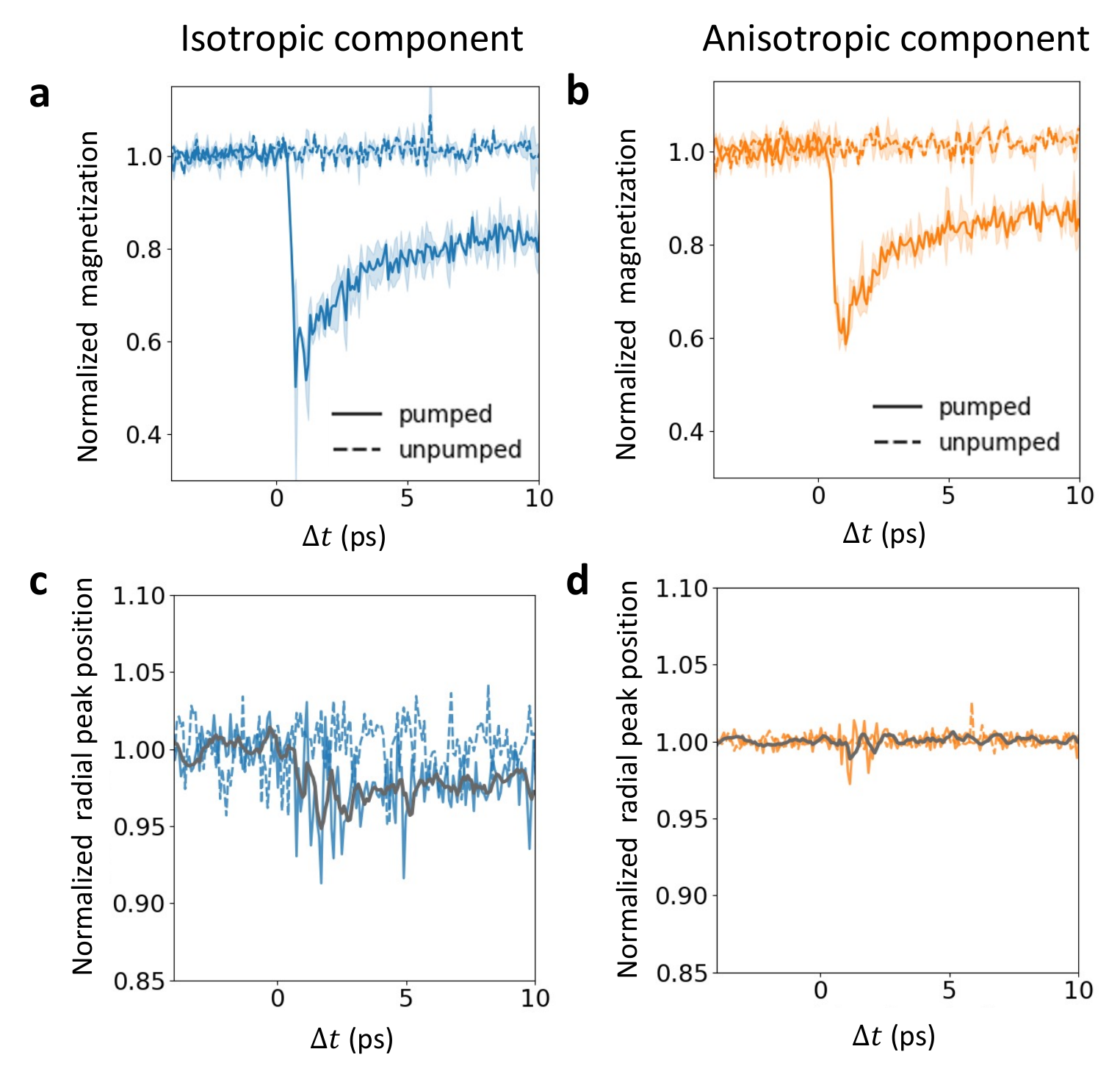}
\caption{\textbf{Ultrafast dynamics of mixed scattering patterns.} Two-dimensional fitting of the full diffraction pattern is used to extract the {time traces of} both the demagnetization and the peak position for the isotropic and anisotropic components. The normalized demagnetization of isotropic and anisotropic components is shown in \textbf{a} and \textbf{b}, respectively. We observe a similar demagnetization in both cases. The temporal evolution of the scattering peak position is shown in \textbf{c} and \textbf{d} for the isotropic and anisotropic components, respectively. As for the samples with a pure symmetry, only the isotropic component exhibits a shift in the peak position. Here we detect a shift of approximately 4\%. The dark line is the five-point moving average of the data. The shaded area corresponds to the standard deviation of the fitted quantities.} 
\label{fig:mixed_states}
\end{center}
\end{figure}

\begin{figure}[t]
\begin{center}
\includegraphics[width=\columnwidth]{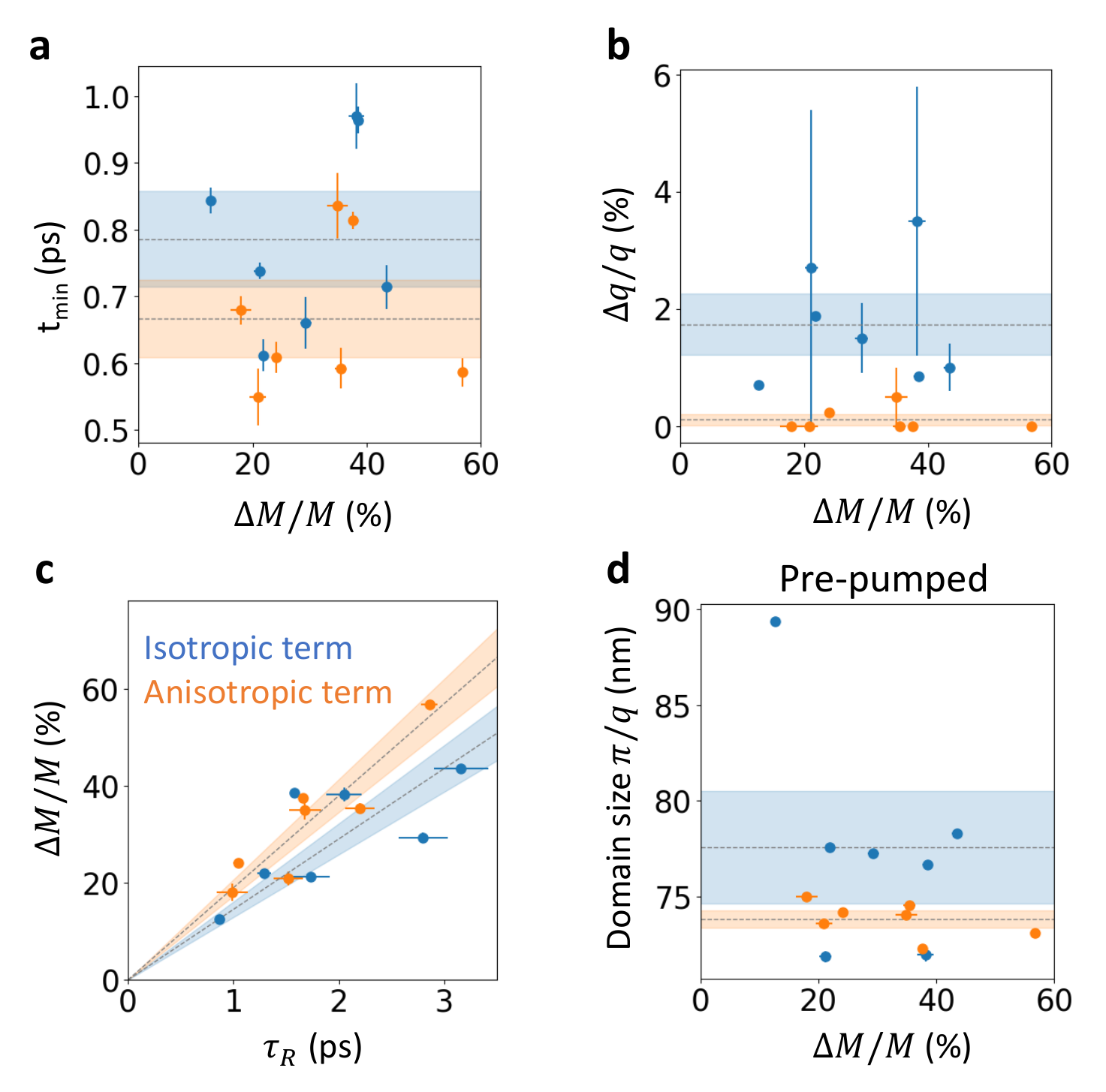}
\caption{\textbf{Key observations}. The extracted \textbf{a} demagnetization time and \textbf{b} maximum change in the {peak-shift} $\Delta q/q$ for the isotropic (blue) and anisotropic (orange) components of the scattering plotted as a function of maximum demagnetization $\Delta M /M$.The dashed line corresponds to the mean value of each set of data and shaded area is the standard deviation. \textbf{c} shows the maximum demagnetization as function of recovery time $\tau_R$. Through linear regression we find that the isotropic and anisotropic components have different recovery speeds. We also plot the corresponding domain size \textbf{d} before the arrival of the pump as a reference state.} 
\label{fig:summary_tmin}
\end{center}
\end{figure}

In addition to the quench, we also detect an ultrafast contraction of the ring radius varying between 1 and 3.5 percent, shown in \cref{fig:pure_states}\textbf{c}. However, we do not observe any change in the peak position of the anisotropic scattering, even at high fluences, as shown in \cref{fig:pure_states}\textbf{d}. These observations are consistent with previous works where a shift in the radial wavevector $\mathbf{q}$ was only observed when measuring isotropic scattering from labyrinth domains~\cite{pfau2012ultrafast, zusin2020ultrafast, leveille2021ultrafast} and no shift reported for anisotropic scattering from stripe domains~\cite{vodungbo2012laserinduced, hennes2020laserinduced}.  We note that the detected shifts (1-4 \%) are smaller than those observed in earlier works (5-6 \%)~\cite{pfau2012ultrafast, zusin2020ultrafast}. We speculate that this difference in our data relative to earlier reports is a consequence of the elevated temperature of our samples due {to} the large effective repetition rate of our instrument (see Supplementary note~\ref{SI:heating}). An elevated temperature before time zero reduces the local magnetic moment such that the net electron-spin scattering that drives the magnetization dynamics is weaker. Regardless, we conclude based on our analysis of the isotropic and anisotropic scattering that the difference in the shifts in $\mathbf{q}$ for stripe and labyrinth domains is not related to varying sample properties or experimental details since we observed this dependence of the shift on domain orientation for the same sample and experiment. We conclude that radial shifts in the diffraction pattern depend on the domain symmetry.

To investigate this further, we turn to the mixed scattering patterns where both isotropic and anisotropic contributions exhibit a similar photon count. A representative example is shown in \cref{fig:mixed_states} {for the time-dependent magnetization, \textbf{a}-\textbf{b}, and peak shift \textbf{c}-\textbf{d}} of the isotropic (blue) and anisotropic (red) components, respectively. As in \cref{fig:pure_states}, the semi-transparent curves represent the unpumped data. We note that the signal-to-noise ratio of the mixed states is lower than that for pure states, but the main features can be recovered from the 2D fits with good accuracy. There are two main observations from \cref{fig:mixed_states}. First, the quench of both symmetry components is similar, estimated to be $38.2\pm1.4$~\% and $34.9\pm1.8$\% for the isotropic and anisotropic components, respectively. Second, there is a distinct shift in the peak position of the isotropic component while no shift can be detected from the anisotropic component. The same trend is observed in all the mixed scattering patterns, as summarized in \cref{table:quench}. Notably, the ultrafast contraction of the isotropic component is consistently larger than any shift in the anisotropic component.

The same behavior was observed for different fluences and runs performed on different days. {The key observations for the quench time $t_\mathrm{min}$, the relative peak position $\Delta q/q$, the recovery time constant $\tau_R$, and the domain size as a function of the quench $\Delta M/M$, are shown in Fig.~\ref{fig:summary_tmin} for both the isotropic (blue) and anisotropic (orange) components.} First of all, we note that the quench time (panel \textbf{a}) does not exhibit a well-defined dependence on $\Delta M/M$ nor a significant dependence on the symmetry component. The horizontal dashed line represents the mean of the data and the shade its standard deviation. Magnetic quench occurs in textured magnetic materials due to both the increase of the magnon population and domain-wall broadening. Therefore, the similar time to quench strongly suggests that these two effects are independent of symmetry. However, there is a clear difference in {peak} shift $\Delta q /q$ between symmetry components, panel \textbf{b}. Indeed, the shift for the anisotropic component (orange) is practically zero. Deviations from zero captured by the fitting procedure are argued to be a consequence of coupling between both symmetry components via their spatial distribution. We note that {there} are two data points that exhibit large error bars for the shift of the symmetric component (blue), indicating that the fits could not be obtained with sufficient confidence to be conclusive. These are correlated with smaller domain sizes at equilibrium (``pre-pumped''), shown in panel \textbf{d}, and concomitant larger linewidths (not shown but reported in Tables~\ref{table:fitted_params_iso} and ~\ref{table:fitted_params_aniso}). This combination resulted in a weaker signal that directly affects our efforts to detect {a peak} shift. Regardless, the data shown in panel \textbf{b} supports the claim that the contraction of the isotropic ring is independent of domain-wall broadening but instead depends on the equilibrium domain symmetry. While the mechanism driving the wavevector shift is currently unknown, our data clearly show that the spatial symmetry of the domains plays a fundamental role in determining the magnitude of the shift.

A further indication of the role of symmetry is found in the recovery time constant, $\tau_R$. In panel \textbf{c} we plot $\Delta M/M$ vs $\tau_R$ and find an approximately linear dependence. The fitted slope is found to differ between the symmetry components, $14.5\pm1.6$~THz for the isotropic component and $19.0\pm1.7$~THz for the anisotropic component{, setting a characteristic recovery speed}. A fluence-dependent recovery time constant was observed previously in Ni undergoing a phase transition to a paramagnetic state~\cite{tengdin2018critical} at the top of the sample. Here, the fluence-dependence recovery time constant does not exhibit a critical behaviour, but suggests that it is dominated by pump-induced disorder. In addition, the longer recovery of the isotropic component suggests that this symmetry is more strongly disordered due to quench.

The extracted characteristic {recovery} speed also provides insights into possible mechanisms driving the recovery. Under the assumption that superdiffusive transport dominates the magnetization quench at a fraction of {picoseconds}~\cite{battiato2010superdiffusive,pfau2012ultrafast}, we turn to electron diffusion for recovery timescales on the picosecond scale. Estimating the diffusion constant of $D\approx1$~cm$^{-2}$/s for our transition metal stack, we find that electronic heat diffusion { along the sample’s plane} has a very long characteristic timescale on the microseconds given the large size of the pump spot, whereas  electronic heat diffusion through the film thickness has a characteristic timescale of the same order of magnitude that we measured, 1-3~ps. Given the sizeable dependence of conductivity on magnetization in transition metals, we propose that the linear dependence of the recovery time on quench amplitude in Fig.~\ref{fig:summary_tmin}\textbf{c} is the result of the concomitant dependence of the electronic diffusion constant on magnetization, with a reduced diffusion constant when the magnetization approaches the Curie temperature. Magnon-mediated heat transport may be excluded as a dominant effect because of typical magnon group velocities under 1~km/s would lead to timescales on the order of 10~ps or greater.

The symmetry dependence on the recovery speed suggests that the magnetization texture has some effect on diffusive electron transport. While domains are preferentially {varying along the sample's plane}, it is already known that domain walls act as scattering sites for electrons~\cite{Levy1997}. Given that the domain wall density per unit area is larger for labyrinth domains relative to oriented stripe domains, it is possible that the increased domain-wall-driven electronic scattering {through the film thickness} in labyrinth domains leads to further reduction in the diffusion constant and, therefore, slower recovery times. 

The results presented here reveal a yet unpredicted dependence of the ultrafast spin dynamics on the nanoscale configuration of magnetic domains. These measurements for the same sample, in the same experimental conditions, and under the same probe spot resolve any controversy as for the legitimate origin of the scattering {peak} shift due to ultrafast pumping. Moreover, our results strongly suggest that this behavior is intrinsically related to the symmetries of the system. The origin of such a shift is still debatable. While a uniform domain expansion can be excluded by impossibly large domain wall motion speeds~\cite{pfau2012ultrafast}, it remains plausible that domains spatially rearrange~\cite{zusin2020ultrafast} or locally demagnetize at different rates. Despite this open question, our results strongly suggest that this shift originates from nonlinear processes that are beyond the assumptions considered for ultrafast demagnetisation processes to date. Surprisingly, these distinct dynamics arise from the symmetry of long-range ordered magnetic domains with sizes ranging between 70 and 90~nm, and extending for several microns, both dimensions longer {than} the typical mean free path of electrons in metallic multilayers~\cite{stohr2006magnetism}. Coupled to the different recovery time constant, our results invite further experimental and theoretical research to clarify the impact of symmetries in the transfer of angular momentum between the electronic and spin degrees of freedom in a far-from-equilibrium setting.

\section*{Methods}
\subsection*{Sample preparation}
The magnetic multilayered samples with the stack layering of (Ta(3~nm)/Cu(5~nm)/[Co$_{90}$Fe$_{10}$(0.25~nm)/Ni(1.35~nm)] \\ x 8 / Co$_{90}$Fe$_{10}$(0.25~nm)/Cu(5~nm)/Ta(3~nm)) were fabricated by sputter-deposition on polycrystalline Si membranes embedded in Si wafer enabling x-ray transmission measurements \cite{shaw2013measurement}. Before the beamtime, MFM measurements showed presence of magnetic out-of-plane labyrinth domains with an average size of 110~nm at remanence.

\subsection*{Experimental setup}
{The experiments are performed at the SCS beamline at the EuXFEL~\cite{Tschentscher2017}}. The XFEL generates 25~fs long, linearly polarized X-ray pulses. In this experiment, we use a pulse-to-pulse separation of 18~$\mu$s, with 26 pulses per train, with the 56~kHz trains having a repetition rate of 10~Hz. This effectively results in 260 pulses per second impinging on the samples. While the EuXFEL is capable of a much higher pulse train frequency with even more pulses per train, longer pulse trains with shorter pulse-to-pulse separation resulted in readily apparent sample damage. Even with these condition, the sample is at an elevated temperature during the time-resolved measurements (see Supplementary Note~\ref{SI:heating}). The incoming X-ray intensity $I_0$ is monitored shot-by-shot using a X-ray Gas Monitor (XGM).

The X-ray scattering is collected on the DSSC 2D detector, able to match the repetition rate of the XFEL~\cite{porro2021minisddbased}. The DSSC records data at twice the X-ray repetition rate in order to collect so-called ``dark'' data frames in between pulses for the best background correction~\cite{zhouhagstrommhzrate}. The sample to detector distance is fixed to 3~m.

{The samples are probed resonantly with the X-rays tuned to the L$_3$ absorption edge of Ni (852~eV). The samples are pumped with an amplified infrared Ti:Sapphire laser with central wavelength $\lambda = 800$~nm and 35~fs pulse duration. The X-ray spot size is estimated to be $20\times20$~$\mu$m$^2$, and the pump laser has a Gaussian profile with $40\times40$~$\mu$m$^2$ at full-width half-maximum. Magnetic domains act as a grating for X-rays at a resonant magnetic edge so that their scattering mathematically represents the two-dimensional Fourier transform of the grating~\cite{hellwig2003xray}. By means of X-ray magnetic circular dichroism (XMCD) contrast~\cite{stohr2006magnetism}, the small-angle X-ray scattering (SAXS) is collected on the DSSC 2D detector~\cite{porro2021minisddbased}.}

\subsection*{2D Fitting procedure and parameter estimation}

The fitting function Eq.~\eqref{eq:2Dfit} contains the isotropic and anistropic components of the scattering. {Under the assumption that both scattering patterns arise from an intermixed spatial pattern, there is no coherent interference contributions to the scattering, as elaborated in the Supplementary Note~\ref{sec:SIinterference}. This implies that the domain patterns exhibit a highly varying spatial periodicity that precludes any possibility of long-range phase coherence of the resultant scattering, as observed in the MFM images of \Cref{fig:setup_SAXS_MFM}\textbf{e}-\textbf{g}.}

Based on previous works~\cite{iacocca2019spincurrentmediated,zusin2020ultrafast}, we define the isotropic scattering intensity is as
\begin{equation}
    f_\mathrm{iso}(\mathbf{q}) = \left[ \frac{A_0}{\left(\mathbf{q}-q_0\right)^2/\Gamma_0^2+1}\right]^2
    \label{eq:2Dring}
\end{equation}
where $A_0$ is the amplitude, $q_0$ is the radius {of the isotropic peak position}, and $\Gamma_0$ is the linewidth. 

The anisotropic scattering can be phenomenologically represented by a Fourier series $\sum{A_n\sin^2{(n(\varphi-\theta))}}$. The intensity is thus defined to second order as
\begin{equation}
f_\mathrm{aniso}(\mathbf{q}, \varphi) = \left[ \frac{ |A_1| \sin^2{(\varphi-\theta)} + A_2\sin^2{(2(\varphi-\theta))}}{\left(\mathbf{q}-q_1\right)^2/\Gamma_1^2+1}\right]^2
\label{eq:2Dlobe}
\end{equation}
This functional form considers that the anisotropic scattering is aligned at an angle $\theta$, has {an anisotropic peak position} $q_1$, and a linewidth $\Gamma_1$. The two amplitude coefficients correspond to the dominant scattering amplitude $A_1$ and a deviation from a sinusoidal azimuthal profile, $A_2$. We find that $A_2$ is typically two orders of magnitude smaller than $A_1$.

Fitting a 2D function requires a detailed and robust protocol: 1) the center of the scattering intensity $\mathbf{q}=0$ is determined at the beginning of each run. Indeed, even a one-pixel offset of the center can generate artefacts such as asymmetries in the radial peak position (see Supplementary Note~\ref{SI:fitbenchmark}). 2) the anisotropic component orientation is determined. 3) The fitting parameters are included subsequently with the goal of determining a good initial guess in an automated way. 4) Fitting of Eq.~\eqref{eq:2Dfit} is performed.

From the fitted parameters, we focus on the magnetization quench and the radial {peak} position of each component. The {average magnetization for each symmetry component} is proportional to the amplitudes {$A_0$ and $A_1$, insofar as $A_2\ll A_1$ for the anisotropic component. The peak position} wavenumber is directly obtained from the {fitted parameters $q_0$ and $q_1$}.

The parameters extracted from all available data sets are presented in the Supplementary Note~\ref{SI:demag}.

\subsection*{Functional form of quench amplitude}
The ultrafast evolution of the magnetization is fitted using
\begin{equation}
\label{eq:quenchfunction}
   f_\mathrm{quench} = 1 +Ae^{-(t-t_0)/\tau_m}-Be^{-(t-t_0)/\tau_R}+ B-A
\end{equation}
where $t_0$ is the time zero of the dynamics,  $\tau_m$ is the quench constant, $\tau_R$ is the recovery constant, $A$ and $B$ are dimensionless constants related to the quench and recovery amplitudes. A more complete form of this equation was derived in Ref.~\cite{unikandanunni2021anisotropic}. Here, we use a simplified form that disregards the longer algebraic recovery constant that could not be fitted accurately within the 20~ps traces.

From this equation we can obtain the quench time $t_\mathrm{min}$ and the maximum quench of the magnetization {$\Delta M /M$} as derived parameters from the fitted variables. Further details on how to obtain such variables and their error are given in Supplementary Note~\ref{SI:demag}. 

\subsection*{Categorizing scattering patterns on photon count}

To categorize our data, we use the following ratio of scattered photons
\begin{equation}
    \label{eq:photon}
    \text{ratio}=\frac{\text{isotropic photons - anisotropic photons}}{\text{total photons}}
\end{equation}
Considering that the background always contributes with a finite amount of photons, isotropic scattering has a ratio close to 1 while anisotropic scattering has a ratio close to $-1$. For a ratio between -0.3 and 0.9 we categorize the scattering pattern as a mixed state. For the categorization of our data, see Supplementary Note~\ref{SI:photoncount}. 

\section*{Acknowledgments}
The authors acknowledge the European XFEL in Schenefeld, Germany, for provision of X-ray free-electron laser beamtime at Scientific Instrument SCS and thank the instrument group and facility staff for their assistance. The authors thank Andrea Castoldi (DSSC consortium) for contributing to the generation of DSSC gain files and Carsten Deiter for the confocal microscopy images. {D.T., and H.A.D. acknowledge support from the Swedish Research Council (VR), Grant 2018-04918. A.H.R. acknowledges support from the US Department of Energy, Office of Science, Office of Basic Energy Sciences under Contract No. DE-AC02-76SF00515.} N.Z.H. and S.B. acknowledge support from the European Research Council, Starting Grant 715452 MAGNETIC-SPEED-LIMIT. R.J., M. and R.K. acknowledge support from AFOSR Grant. No. FA9550-19-1-0019.

\section*{Author Contributions}
N.Z.H., R.J., M., D.T., J.B., E.L., B.S., J.S., N.G., B.V.K., R.G., L.M., L.L.G., A.S., A.S., G.M., H.D., H.N., J.M.S., M.W.K., E.J., E.E.F., R.K., S.B., T.J.S., and E.I. performed the experiment. {M.A. and }J.M.S. deposited and characterized the samples. N.Z.H., R.J., M., M.K., R.K., T.J.S., and E.I. analyzed the data. All authors discussed the results and contributed to writing the paper.

\section*{Competing Interests statement}

The authors declare no competing interests. 

\section*{Corresponding authors}
\noindent Correspondence to\\ eiacocca@uccs.edu,\\thomas.silva@nist.gov, and \\ stefano.bonetti@fysik.su.se

%

\onecolumngrid
\clearpage

\Large{\textbf{Supplementary material: Nanoscale magnetic configuration dependent ultrafast spin dynamics}}
\\

\large
Nanna Zhou Hagström et al.
\newpage

\setcounter{section}{0}
\renewcommand{\thesection}{SI \arabic{section}}

\setcounter{figure}{0}
\renewcommand{\thefigure}{SI \arabic{figure}}

\setcounter{table}{0}
\renewcommand{\thetable}{SI \Roman{table}}

\setcounter{equation}{0}
\renewcommand{\theequation}{SI \arabic{equation}}

\section{Strain in multilayers}
\label{sec:SIstrain}

\begin{figure}[ht]
\begin{center}
\includegraphics[width=0.5\textwidth]{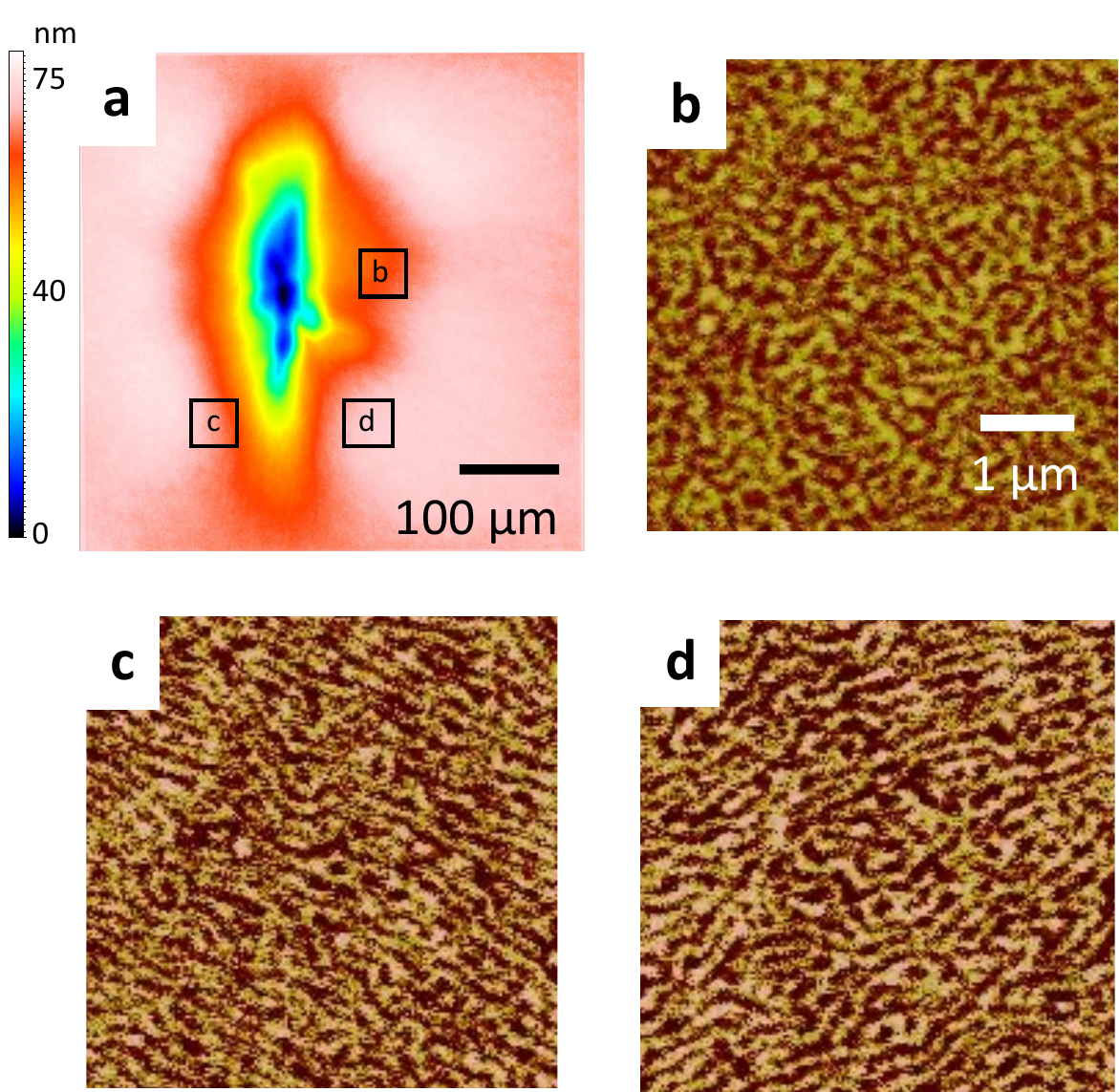}
\caption{\textbf{Strain-induced domain anisotropy} \textbf{a} Confocal microscope image of a membrane after X-ray and IR laser exposure. MFM measurements were performed on the membrane after the beamtime. $5 \times 5$~$\mu$m$^2$ images taken at different areas of the membranes are showing \textbf{b} labyrinth domains, \textbf{c} partially oriented domains and \textbf{d} partially oriented along the opposite direction.}
\label{fig:strain}
\end{center}
\end{figure}

The magnetic multilayers were deposited on polycristalline Si membranes. The measured membranes were characterized post-beamtime using confocal microscopy and magnetic force microscopy (MFM). From the confocal microscopy image \textbf{a} we observe that the X-rays and IR laser left \textit{imprints} of up to 75~nm in the membranes. This strain-induced distortion of the sample would explain why we observed anisotropy in the scattering pattern. Indeed the MFM measurements confirm that the domains were orientated differently throughout the sample with some areas presenting \textbf{b} labyrinth domains while others \textbf{c}-\textbf{d} showed partially oriented domains with different orientation.         

\section{Loss of coherent interference in the scattering pattern}
\label{sec:SIinterference}

We demonstrate that the fitting function can be approximated by the addition of two intensities without any cross-correlation. For simplicity, we consider a 1D model with regions of width $a$, each with a phase $\phi$ and a periodicity $q_n$. The 1D function $f_1$ can be expressed as an infinite sum of such regions, given by
\begin{equation}
\label{eq:sc5}
   f_1(x) = \sum_n^\infty{\sin(q_nx+\phi_n)\Pi((x+n/2)/a)},
\end{equation}
where $\Pi(x/a)$ is the pulse function of width $a$.

By use of standard Fourier transform properties, we obtain the Fourier transform $F_1(q)=\mathcal{F}\{f_1(x)\}$ as
\begin{eqnarray}
\label{eq:sc6}
   F_1(q) = \frac{a}{2}\sum_n^\infty&&\left[\cos(\phi_n)e^{i\pi/2}+\sin(\phi_0)\right]e^{ina(q+q_n)/2}\mathrm{sinc}\left(\frac{a(q+q_n)}{2\pi}\right)\nonumber\\&&+\left[\cos(\phi_n)e^{-i\pi/2}+\sin(\phi_0)\right]e^{ina(q-q_n)/2}\mathrm{sinc}\left(\frac{a(q-q_n)}{2\pi}\right).
\end{eqnarray}

Therefore, the intensity is
\begin{eqnarray}
\label{eq:sc7}
   |F_1(q)|^2 &=& \frac{a^2}{4}\sum_n^\infty\mathrm{sinc}^2\left(\frac{a(q+q_n)}{2\pi}\right)+\mathrm{sinc}^2\left(\frac{a(q-q_n)}{2\pi}\right)\nonumber\\&+&\frac{2a^2}{4}\sum_{n,m}^\infty\Big[\cos\left(\phi_m-\phi_n-\frac{a}{2}(nq_n-mq_m)\right)\nonumber\\&&\mathrm{sinc}\left(\frac{a(q+q_n)}{2\pi}\right)\mathrm{sinc}\left(\frac{a(q+q_m)}{2\pi}\right)\Big]
\end{eqnarray}

The third term in Eq.~\eqref{eq:sc7} represents a cross-term from the overlap of the sinc functions. If we assume that the periods between neighboring regions are similar, $q_n\approx q_m$, then only those regions where $n\approx m$ will contribute to the spectrum. If this were not the case, then overlap between the sinc functions would be small, and the cross-term would be negligible. It follows that the cross-term is dominantly proportional to $\cos\left(\phi_m-\phi_n\right)$.

If the phase between neighboring regions is randomized, then the infinite sum over $\phi_m-\phi_n$ will span every phase between 0 and $2\pi$. Therefore the cross-term is zero. This situation is equivalent to consider that each region has a finite correlation length.

This toy model illustrates that the cross-term can be neglected for the ``mixed'' state, where the spatial periodicity is randomized.

\section{2D fitting procedures and quality estimation}
\label{SI:fitbenchmark}

\begin{figure}[ht]
\begin{center}
\includegraphics[width=\textwidth]{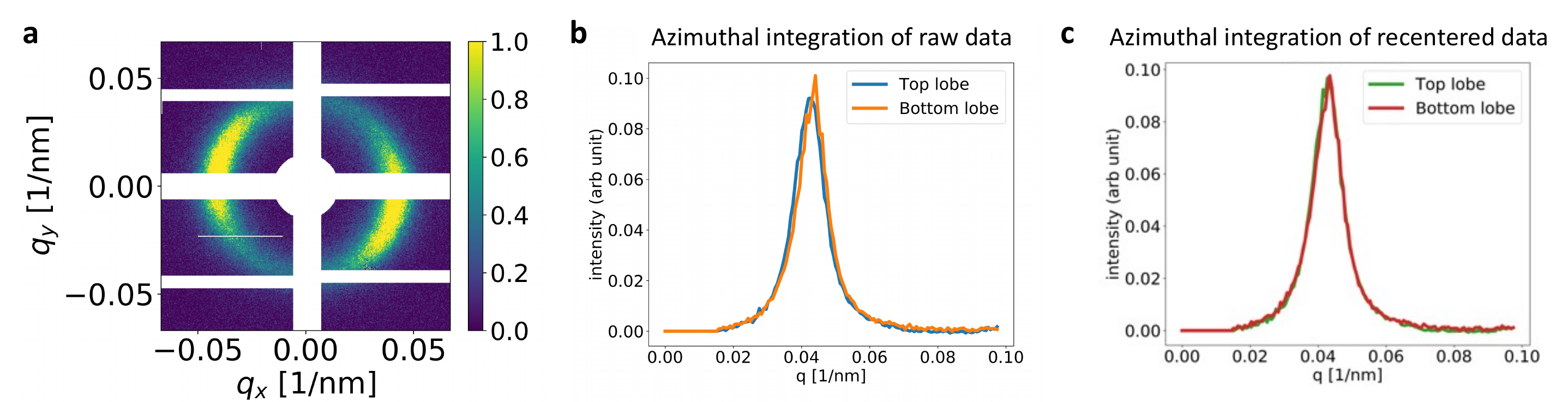}
\caption{\textbf{Centering the two dimensional fit} \textbf{a} The raw data collected on the DSSC. The azimuthal integration over the top left and bottom right lobes are plotted for the raw image \textbf{b} and for the re-centered image \textbf{c}. The re-centered image has been shifted by  +0.7 pixel in x and +1.8 pixels in y.}
\label{fig:SI_centering}
\end{center}
\end{figure}

We highlight the importance of fitting the center of the scattering data instead of relying on the center of the raw image from the DSSC. For example, we consider the scattering data shown in \Cref{fig:SI_centering}\textbf{a}. Taking $\mathbf{q}=0$ as the center of the scattering intensity, we azimuthally integrate the left and right modules of the scattering. As a result, we observe in \Cref{fig:SI_centering}\textbf{b} that the azimuthal profile of the left and right features do not fully overlap at the same wavenumber. This is not physical because there cannot be a preferential scattering angle from a randomized domain pattern. By including the center of the scattering as a fitting parameter, we are able to determine the true $\mathbf{q}=0$ as a function of pixels in the DSSC detector. The azimuthal profile of the features then overlap perfectly as shown in \cref{fig:SI_centering}\textbf{c}. We find that the offset is typically within 2 pixels, which is less than 0.5~mm. We keep the center as a floating parameter when fitting the time-resolved measurements. 

\begin{figure}[ht]
\begin{center}
\includegraphics[width=\textwidth]{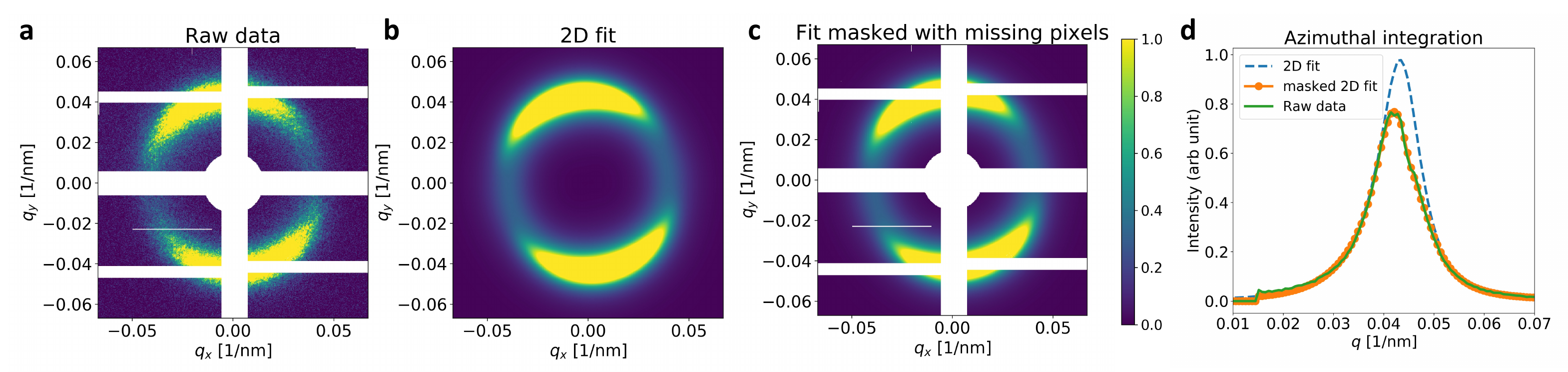}
\caption{\textbf{Effect of missing pixels on azimuthally integrated data.} \textbf{a} The raw scattering collected on the DSSC. \textbf{b} The result of the 2D fit. \textbf{c} 2D fit masked with the missing pixels of the detector. \textbf{d} Azimuthally integrated scattering of the raw data (solid green curve), 2D fit (dashed blue curve) and 2D fit masked with DSSC inactive areas (dotted orange curve). The 2D fit only matches the raw data once it has been masked with the DSSC inactive areas.}
\label{fig:missing_pixels}
\end{center}
\end{figure}

The inactive areas of the DSSC sometimes cover substantial parts of the diffraction pattern as seen in \cref{fig:missing_pixels}\textbf{a}. The 2D fit extrapolates the intensity in areas with missing pixels and allows us to reconstruct the full scattering, as shown in \cref{fig:missing_pixels}\textbf{b}. 

A natural question is whether this reconstruction is accurate and physically meaningful. A simple test is to azimuthally average the data as shown in \cref{fig:missing_pixels}\textbf{d}. We observe a mismatch between the raw data (green solid curve) and the fit (dotted blue curve). In particular, the raw data exhibits a shoulder at $q\approx0.05$~nm$^{-1}$ that suggests a bimodal distribution. However, if we mask the 2D fit with the inactive areas of the DSSC  as shown in \cref{fig:missing_pixels}\textbf{c}, the azimuthal integration of this data (orange dotted curve) overlaps with the raw data. This shows that the intensity mismatch, asymmetry, and shoulder in the raw data are only due to the inactive areas of the DSSC and indicates that great care must be taken when extracting information from azimuthally averaged or integrated data from a modular DSSC. This artefacts can point to inexistent physics in the ultrafast regime.

\begin{figure}[ht]
\begin{center}
\includegraphics[width=\textwidth]{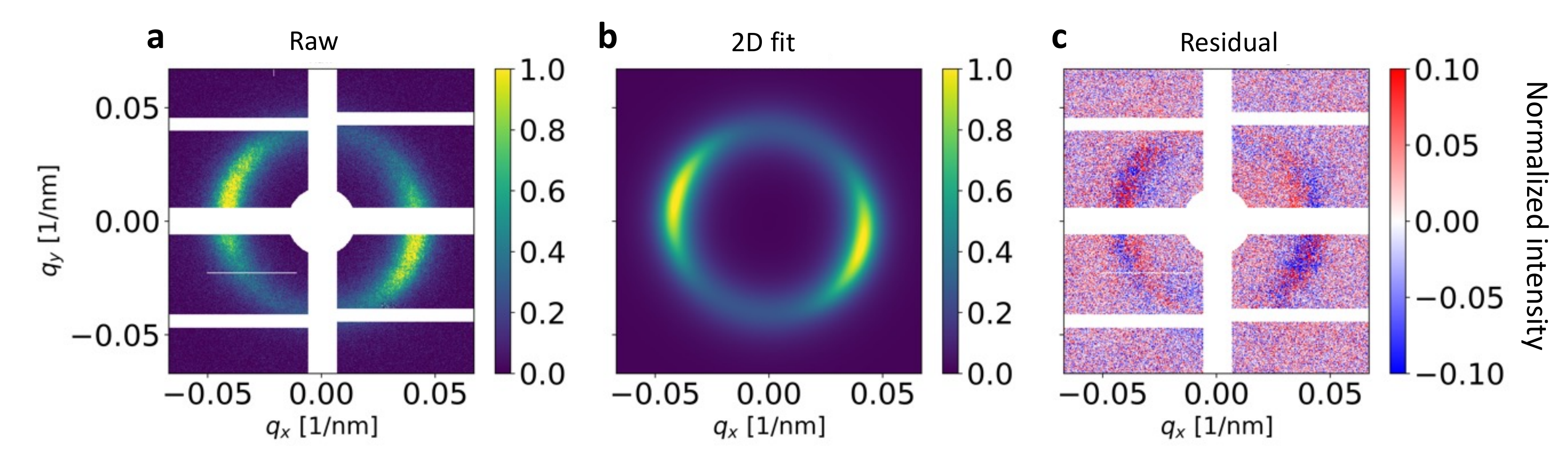}
\caption{\textbf{2D fit and residual.} \textbf{a} A raw data collected on the DSSC. \textbf{b} Two dimensional fit of \textbf{a}. The resulting residual (raw - fit) is shown in panel \textbf{c}. Note that the intensity scale of the residual is one order of magnitude smaller than that of the raw scattering pattern.}
\label{fig:SI_2Dresidual}
\end{center}
\end{figure}

Finally, the quality of the fit is determined from the residual between the raw data and the fit. In \cref{fig:SI_2Dresidual} we shown a example scattering, its fit, and the its residual. There is a clear finite residue at the location of the ring that indicates that our phenomenological function fails to correctly account for the detailed scattering profile. However, the residual is on the order of 10~\%. In addition, the residual fluctuates rapidly, which can be also a consequence of shot noise, proportional to the intensity, and the speckle pattern. We then conclude that our fits are sufficiently accurate to extract physically meaningful information.

\begin{figure}[ht]
\begin{center}
\includegraphics[width=0.7\textwidth]{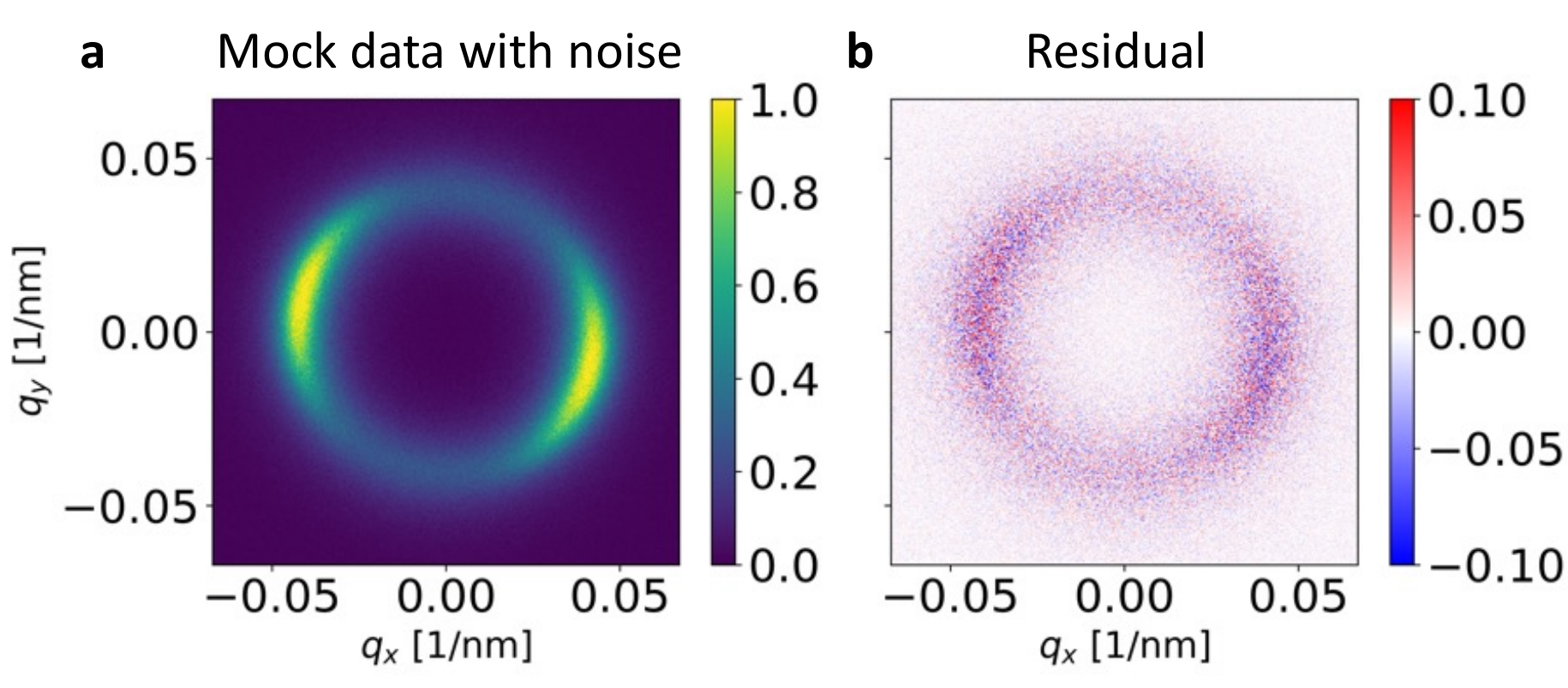}
\caption{\textbf{2D fit trial.} \textbf{a} A mock scattering pattern. \textbf{b} The resulting residual of the 2D fit. Note that the intensity scale of the residual is one order of magnitude smaller than that of the raw scattering pattern.}
\label{fig:SI_2d_fit_trial}
\end{center}
\end{figure}

We further validate our results by showing that the 2D fit is able to track small changes in the anisotropic peak position $q_1$. We generate a mock scattering pattern with added shot noise (\cref{fig:SI_2d_fit_trial}\textbf{a}) and a shift in $q_1$ of approximately $1$\%. The shift in $q_1$ is recovered by the fitting procedure within error. The parameters are given in \cref{table:fit_trial_params}. The residual \textbf{b} shows the difference between the mock data and the fit. There is a clear residue with fluctuating values due to the shot noise of the mock data.

\begin{table}[t]
\centering
\ra{1.1}
\begin{adjustbox}{width=0.7\textwidth}
\begin{tabular}{l l l l l}
\toprule[1pt]
\text{2D fit parameters} & \text{Parameters of mock data} & \text{Initial guess}       & \multicolumn{2}{c}{Fitted parameters}   \\ \midrule[1pt]
B                 & 6.66e-6    & 6.66e-6   & 6.62   & 0.03e-6       \\
q$_0$             & 4.0183e-2  & 4.0183e-2 & 4.0188 & 0.0005e-2   \\
$\Gamma_0$        & 1.448 e-2  & 1.448e-2  & 1.451  & 0.001e-2      \\
A$_0$             & 1.6052e-2  & 1.6052e-2 & 1.6050 & 0.0005e-2   \\
q$_1$             & 4.2636e-2  & 4.2136e-2 & 4.2635 & 0.0002e-2    \\
$\Gamma_1$        & 0.7011e-2  & 0.7011e-2 & 0.6991 & 0.0004e-2    \\
$|\mathrm{A}_1|$  & 2.8026e-2  & 2.8026e-2 & 2.8026 & 0.0007e-2    \\
$|\mathrm{A}_2|$  & 0.111e-2   & 0.111e-2  & 0.111  & 0.001e-2      \\
$\varphi$         & 0.244      & 0.244     & 0.245  & 0.002                    \\ 
\bottomrule[1pt]
\end{tabular}
\end{adjustbox}
\caption{\textbf{Parameters of the phenomenological 2D fit function.} We give the parameters used to generate the mock data before adding shot noise (\cref{fig:SI_2d_fit_trial}\textbf{a}). In the initial guess, the anisotropic scattering radius q$_1$ is slightly shifted. This change is recovered by the 2D fit as shown in the last column.}
\label{table:fit_trial_params}
\end{table}

\section{Pulse resolved photon count in the scattering data}
\label{SI:photoncount}

\begin{figure}[ht]
\begin{center}
\includegraphics[width=\textwidth]{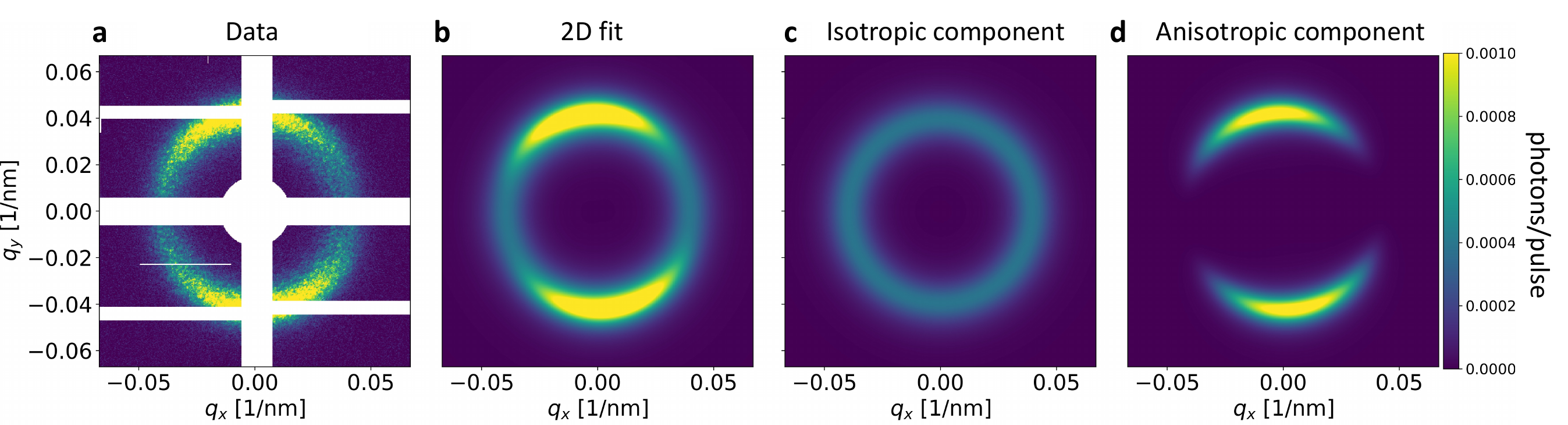}
\caption{\textbf{Two dimensional fit of the scattering data} \textbf{a} The raw data collected on the DSSC. The resulting two dimensional fit (b) can be separated into ring component \textbf{c} and lobe component \textbf{d}. The intensity colorbar is shared between all four panels and can be interpreted as photons/pixel/pulse.}
\label{fig:SI_photon count}
\end{center}
\end{figure}

The DSSC is a soft X-ray detector with single-photon sensitivity that is able to match the repetition rate of the FEL \cite{porro2021minisddbased}. The resulting scattering image (\cref{fig:SI_photon count}\textbf{a}) is the average of all frames collected within a run. After multiplying by the gain of the detector, the intensity can be directly related to the average number of photons hitting each pixel per pulse. In our experiment, the gain was set to 0.5~photon/bin. 

The two-dimensional fit allows us to extrapolate the intensity in the areas with missing pixels as well as fully separate the isotropic from the anisotropic components (see main text). It is therefore possible to calculate the average number of photons per X-ray pulse contributing to each term by summing the intensity in the resulting image. In the example of \cref{fig:SI_photon count}, an average of 17.5 photons per pulse hit the detector where 9.5 photons belong to the isotropic contribution and 7.5 photons to the anisotropic contribution. The background contributes with an average of 0.5 photon/pulse. With 26 pulses per train at a train repetition rate of 10~Hz, we get 4550 photons per second in the full scattering with 2470 photons in the isotropic component, 1950 photons in the anisotropic component, and 130 photons in the background.  \cref{table:photon count} summarizes the photon count from all the pump-probe measurements.

\begin{table}[t]
\centering
\ra{1.1}
\begin{adjustbox}{width=\textwidth}
\begin{tabular}{lcccccccc} \toprule[1pt]
    & \multicolumn{4}{c}{Total photons per pulse} & \multicolumn{4}{c}{Total photons per second}  \\ \cmidrule(lr){2-5}\cmidrule(lr){6-9}
Fluence & Full scattering & Anisotropic & Isotropic  & Background & Full scattering & Anisotropic & Isotropic  & Background \\ \midrule[1pt]
20~mJ cm$^{-2}$ FO (day 3) & 13.55  & 4.35  & 7.85  & 1.35  & 3523  & 1131 & 2041 & 351 \\ 
15~mJ cm$^{-2}$ FO (day 3) & 14.30  & 4.55  & 8.4   & 1.35  & 3718  & 1183 & 2184 & 351 \\ 
25~mJ cm$^{-2}$ FO (day 4) & 16.75  & 3.35  & 10.15 & 3.25  & 4355  & 871  & 2639 & 845 \\ 
15~mJ cm$^{-2}$ FO (day 4) & 21.40  & 3.25  & 12.1  & 6.05  & 5564  & 845  & 3146 & 1573\\ 
15~mJ cm$^{-2}$ FO (day 5) & 18.38  & 12.21 & 5.29  & 0.88  & 4778  & 3175 & 1375 & 228 \\
15~mJ cm$^{-2}$ PO (day 5) & 18.82  & 8.43  & 9.49  & 0.9   & 4891  & 2190 & 2467 & 234 \\
15~mJ cm$^{-2}$ PO (day 5) & 22.21  & 0     & 21.18 & 1.03  & 5772  & 0    & 5507 & 267 \\
10~mJ cm$^{-2}$ FO (day 5) & 20     & 0     & 18.35 & 1.65  & 5200  & 0    & 4771 & 429 \\
10~mJ cm$^{-2}$ FO (day 5) & 18.95  & 13.34 & 4.84  & 0.77  & 4927  & 3468 & 1258 & 201 \\  \bottomrule[1pt]
\end{tabular}
\end{adjustbox}
\caption{\textbf{Average photon count quantified from the 2D fit} The average photon count is quantified per pulse and per second. For all the runs the XFEL delivered 26 pulses per train with a train repetition rate of 10~Hz. The average photon count in the full scattering is decomposed into the anisotropic and isotropic components as well as the uniform non-magnetic background. }
\label{table:photon count}
\end{table}

\begin{figure}[ht]
\begin{center}
\includegraphics[width=0.4\textwidth]{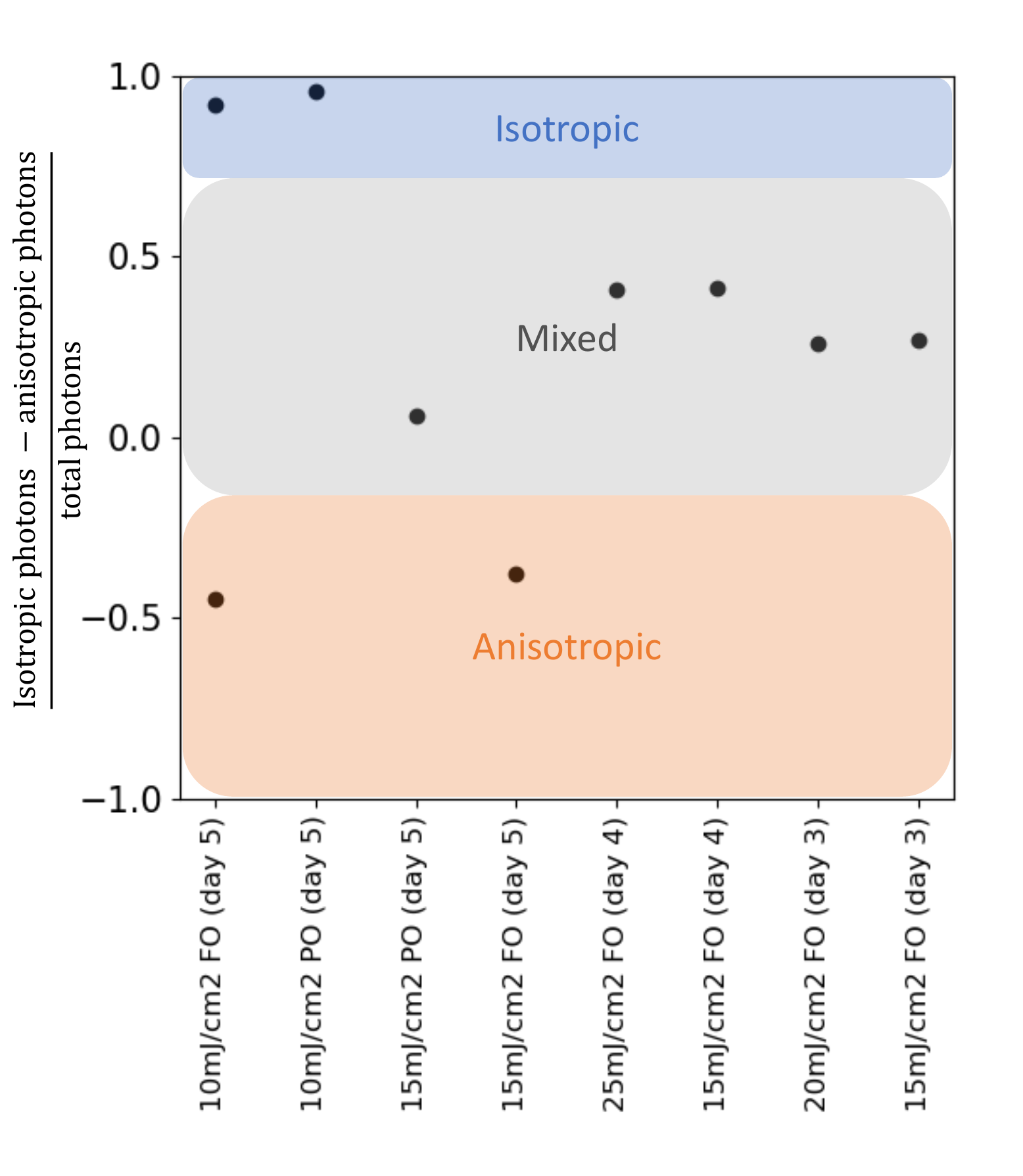}
\caption{\textbf{Distribution of data according to categorization} The ratio is computed according to Eq.~\eqref{eq:photon}. Isotropic scattering has a ratio close to 1 while anisotropic scattering has a ratio close to -1. For a ratio between -0.3 and 0.9 we categorize the scattering pattern as mixed.}
\label{fig:SI_state_ratio}
\end{center}
\end{figure}

In the Methods section we explain how we categorize our data according to Eq.~\eqref{eq:photon}. \Cref{fig:SI_state_ratio} shows the distribution of ratios for our data. We can clearly distinguish three groups that correspond to the three types of scattering that we observe. Due to the non-zero background, isotropic scattering has a ratio close to 1 while anisotropic scattering has a ratio close to -1. We note that, because the anisotropic photons are more localized in space, fewer photons are required to observe anisotropic scattering, while more photons are needed in order to clearly observe isotropic scattering above the noise.   

\section{Fitting the demagnetization constants}
\label{SI:demag}

The fitting procedure for extracting demagnetization values and time constants is explained in detail by Unikandanunni \textit{et al.}. We derive the equation from  \cite{unikandanunni2021anisotropic}:
\begin{equation}
\label{eq:1}
    \frac{\Delta M}{M_0} = (\frac{A_1 \tau_R - A_2\tau_m}{\tau_R-\tau_m}e^{-(t-t_0)/\tau_m} - \frac{\tau_R(A_1-A_2)}{\tau_R-\tau_m} e^{-(t-t_0)/\tau_R} - \frac{A_2}{\sqrt{(t-t_0)/\tau_{R2}+1}}) \circledast \Gamma(t)
\end{equation}
where $\tau_m$ is the quench constant, $\tau_R$ is the recovery constant, $A_1$ and $A_2$ are dimensionless constants related to the quench and recovery amplitudes. We fit $t_0$, the time zero of the dynamics, in order to account for drifts of the pump arrival. The expression within parentheses is convoluted with a Gaussian $\Gamma(t)$ to account for the finite duration of the X-ray and IR pulses.     

We disregard the second, algebraic recovery time Unikandanunni \textit{et al.} used in their work since we did not collected enough data at longer times to properly fit this value. Therefore, we use a simpler expression given by
\begin{equation}
\label{eq:4}
   f_\mathrm{quench} = 1 +Ae^{-(t-t_0)/\tau_m}-Be^{-(t-t_0)/\tau_R}+ B-A,
\end{equation}
where the new coefficients, $A$ and $B$, are related to the coefficients in Eq.~\eqref{eq:1} by
\begin{subequations}
\label{eq:5}
\begin{eqnarray}
   A_1 &=& A-B\frac{\tau_m}{\tau_R},\\
	 A_2 &=& A-B.
\end{eqnarray}
\end{subequations}

From this equation, we can analytically obtain the quench time by finding the minimum of $f_\mathrm{quench}$, given by
\begin{equation}
\label{eq:6}
   t_\mathrm{min} = t_0 - \frac{\tau_m\tau_R}{\tau_R-\tau_m}\ln\left(\frac{B\tau_m}{A\tau_R}\right).
\end{equation}

It follows that the error $\delta t_\mathrm{min}$ can be computed by standard propagation of uncertainty for each fitted variable. These are:
\begin{subequations}
\label{eq:7}
\begin{eqnarray}
  \frac{\partial t_\mathrm{min}}{\partial t_0} &=& \delta t_0,\\
	\frac{\partial t_\mathrm{min}}{\partial A} &=& \frac{\tau_m\tau_R}{\tau_R-\tau_m}\frac{\delta A}{A}\approx\frac{\tau_m}{A}\delta A,\\
	\frac{\partial t_\mathrm{min}}{\partial B} &=& \frac{\tau_m\tau_R}{\tau_R-\tau_m}\frac{\delta B}{B}\approx\frac{\tau_m}{B}\delta B,\\
	\frac{\partial t_\mathrm{min}}{\partial \tau_m} &=& \frac{\tau_R}{\tau_R-\tau_m}\left[1+\frac{\tau_R}{\tau_R-\tau_m}\ln\left(\frac{B\tau_m}{A\tau_R}\right)\right]\delta \tau_m\approx\left[1+\ln{\frac{B\tau_m}{A\tau_R}}\right]\delta\tau_m,\\
	\frac{\partial t_\mathrm{min}}{\partial \tau_R} &=& \frac{\tau_m}{\tau_R-\tau_m}\left[1+\frac{\tau_m}{\tau_R-\tau_m}\ln\left(\frac{B\tau_m}{A\tau_R}\right)\right]\delta \tau_R\approx\frac{\tau_m}{\tau_R}\delta\tau_R.
\end{eqnarray}
\end{subequations}

The approximations consider that $\tau_m\ll\tau_R$, but this leads to an infinite error in $\tau_m$, so the approximation must be taken with care.

The quench is then $f_\mathrm{quench}$ evaluated at $t_\mathrm{min}$,
\begin{equation}
\label{eq:8}
   A_q = A\left[1 -\left(\frac{B\tau_m}{A\tau_R}\right)^{\tau_R/(\tau_R-\tau_m)}\right]-B\left[1 -\left(\frac{B\tau_m}{A\tau_R}\right)^{\tau_m/(\tau_R-\tau_m)}\right],
\end{equation}
and the error $\delta A_q$ can be obtained from propagation of uncertainty of the following quantities
\begin{subequations}
\label{eq:9}
\begin{eqnarray}
  \frac{\partial A_q}{\partial A} &=& \left[1-\left(\frac{B\tau_m}{A\tau_R}\right)^{\tau_R/(\tau_R-\tau_m)}\right]\delta A\approx\delta A,\\
	\frac{\partial A_q}{\partial B} &=& \left[1-\left(\frac{B\tau_m}{A\tau_R}\right)^{\tau_m/(\tau_R-\tau_m)}\right]\delta B,\\
	\frac{\partial A_q}{\partial \tau_R} &=& \frac{1}{\tau_R(\tau_R-\tau_m)^2}\Big[A\tau_R\left(\frac{B\tau_m}{A\tau_R}\right)^{\tau_R/(\tau_R-\tau_m)}\left(\tau_R-\tau_m+\tau_m\ln\left(\frac{B\tau_m}{A\tau_R}\right)\right)\nonumber\\
	&-&B\tau_m\left(\frac{B\tau_m}{A\tau_R}\right)^{\tau_m/(\tau_R-\tau_m)}\left(\tau_R-\tau_m+\tau_R\ln\left(\frac{B\tau_m}{A\tau_R}\right)\right)\Big]\delta \tau_R,\approx\frac{B\tau_m}{\tau_R^2}\delta\tau_R\\
	\frac{\partial A_q}{\partial \tau_m} &=& \frac{1}{\tau_m(\tau_R-\tau_m)^2}\Big[B\tau_m\left(\frac{B\tau_m}{A\tau_R}\right)^{\tau_m/(\tau_R-\tau_m)}\left(\tau_R-\tau_m+\tau_R\ln\left(\frac{B\tau_m}{A\tau_R}\right)\right)\nonumber\\
	&-&A\tau_R\left(\frac{B\tau_m}{A\tau_R}\right)^{\tau_R/(\tau_R-\tau_m)}\left(\tau_R-\tau_m+\tau_m\ln\left(\frac{B\tau_m}{A\tau_R}\right)\right)\Big]\delta \tau_m\approx\frac{B}{\tau_R}\delta\tau_m.
\end{eqnarray}
\end{subequations}

\cref{table:fitted_params_iso} and \cref{table:fitted_params_aniso} show the extracted fitting parameters discussed in our work for all measurements performed during the experiment. 

\begin{table}[h]
\begin{adjustbox}{width=\textwidth}
\begin{tabular}{ccccccccc} \toprule[1pt]
 \multicolumn{2}{c}{}  & \multicolumn{7}{c}{Isotropic component} \\ \cmidrule(lr){3-9}
  \multicolumn{2}{c}{}  & \multicolumn{2}{c}{Pre-pumped} & \multicolumn{5}{c}{Pumped} \\ \cmidrule(lr){3-4}\cmidrule(lr){5-9}
Fluence & SAXS pattern & size (nm) & $\Gamma $ ($\mu$m$^{-1}$) & $\Delta q/ q$ (\%) & $\Delta M / M$ (\%) & t$_{min}$ (ps) & quench speed (ps$^{-1}$) & $\tau_{R} $ (ps) \\ \midrule 
20~mJ cm$^{-2}$ FO (day 3) & mixed       & 78.30 $\pm$ 0.11 & 1.482 $\pm$ 0.012          & 1.0 $\pm$ 0.4          & 43.5 $\pm$ 0.9 & 0.714 $\pm$ 0.033 & 60.9 $\pm$ 3.1 & 3.16 $\pm$ 0.27  \\ 
15~mJ cm$^{-2}$ FO (day 3) & mixed       & 77.27 $\pm$ 0.14 & 1.679 $\pm$ 0.017          & 1.5 $\pm$ 0.6          & 29.3 $\pm$ 1.0 & 0.660 $\pm$ 0.039 & 44.4 $\pm$ 3.0 & 2.80 $\pm$ 0.23  \\ 
25~mJ cm$^{-2}$ FO (day 4) & mixed       & 71.96 $\pm$ 0.34 & \textbf{2.572 $\pm$ 0.055} & 3.5 $\pm$ 2.3          & 38.2 $\pm$ 1.4 & 0.970 $\pm$ 0.049 & 39.4 $\pm$ 2.5 & 2.05 $\pm$ 0.17  \\ 
15~mJ cm$^{-2}$ FO (day 4) & mixed       & 71.86 $\pm$ 0.28 & \textbf{2.534 $\pm$ 0.045} & 2.7 $\pm$ \textbf{2.7} & 21.2 $\pm$ 1.0 & 0.738 $\pm$ 0.012 & 28.7 $\pm$ 1.4 & 1.73 $\pm$ 0.18  \\ 
15~mJ cm$^{-2}$ PO (day 5) & mixed       & 77.56 $\pm$ 0.04 & 1.086 $\pm$ 0.004          & 1.88 $\pm$ 0.07        & 21.9 $\pm$ 0.7 & 0.610 $\pm$ 0.02  & 35.8 $\pm$1.8  & 1.29 $\pm$ 0.07  \\ 
10~mJ cm$^{-2}$ PO (day 5) & isotropic   & 89.38 $\pm$ 0.02 & 1.249 $\pm$ 0.002          & 0.70 $\pm$ 0.05        & 12.6 $\pm$ 0.5 & 0.840 $\pm$ 0.02  & 14.9 $\pm$ 0.7 & 0.87 $\pm$ 0.05  \\ 
10~mJ cm$^{-2}$ FO (day 5) & isotropic   & 76.66 $\pm$ 0.01 & 0.953 $\pm$ 0.002          & 0.84 $\pm$ 0.06        & 38.5 $\pm$ 0.8 & 0.964 $\pm$ 0.02  & 39.9 $\pm$ 1.2 & 1.57 $\pm$ 0.03  \\ \bottomrule[1pt]
\end{tabular}
\end{adjustbox}
\caption{\textbf{Extracted parameters from the isotropic component of the 2D fit} Measurements are listed by pump fluence and scattering pattern. We distinguish between full overlap (FO) and partial overlap (PO) between probe and probe.  Extracted domain size $\pi/q$  (nm) and linewidth $\Gamma$ (nm$^{-1}$) from the pre-pumped signal, maximum shift in radial peak position (\%), maximum demagnetization (\%), quench time (ps) and quench speed (ps$^{-1}$) and demagnetization recovery time (ps) for all measurements.}
\label{table:fitted_params_iso}
\end{table}

\begin{table}[h]
\begin{adjustbox}{width=\textwidth}
\begin{tabular}{ccccccccc} \toprule[1pt]
 \multicolumn{2}{c}{}  & \multicolumn{7}{c}{Anisotropic component} \\ \cmidrule(lr){3-9}
  \multicolumn{2}{c}{}  & \multicolumn{2}{c}{Pre-pumped} & \multicolumn{5}{c}{Pumped} \\ \cmidrule(lr){3-4}\cmidrule(lr){5-9}
Fluence & SAXS pattern & size (nm) & $\Gamma $ ($\mu$m$^{-1}$) & $\Delta q/ q$ (\%) & $\Delta M / M$ (\%) & t$_{min}$ (ps) & quench speed (ps$^{-1}$) & $\tau_{R} $ (ps)  \\ \midrule 
20~mJ cm$^{-2}$ FO (day 3) & mixed       & 74.55 $\pm$ 0.04 & 0.711 $\pm$ 0.004  & 0                      & 35.4 $\pm$ 1.0 & 0.549 $\pm$ 0.043 & 59.8 $\pm$ 3.5 & 2.25 $\pm$ 0.14  \\ 
15~mJ cm$^{-2}$ FO (day 3) & mixed       & 73.58 $\pm$ 0.04 & 0.715 $\pm$ 0.005  & 0                      & 20.9 $\pm$ 1.4 & 0.549 $\pm$ 0.043 & 38.1 $\pm$ 3.9 & 1.52 $\pm$ 0.14  \\ 
25~mJ cm$^{-2}$ FO (day 4) & mixed       & 74.06 $\pm$ 0.14 & 0.808 $\pm$ 0.016  & \textbf{0.5 $\pm$ 0.5} & 34.9 $\pm$ 1.8 & 0.836 $\pm$ 0.049 & 41.7 $\pm$ 3.3 & 1.68 $\pm$ 0.15  \\ 
15~mJ cm$^{-2}$ FO (day 4) & mixed       & 74.98 $\pm$ 0.14 & 0.794 $\pm$ 0.015  & 0                      & 18.0 $\pm$ 1.8 & 0.679 $\pm$ 0.021 & 26.5 $\pm$ 2.8 & 0.99 $\pm$ 0.15  \\ 
15~mJ cm$^{-2}$ FO (day 5) & anisotropic & 73.08 $\pm$ 0.01 & 0.729 $\pm$ 0.001  & 0                      & 56.7 $\pm$ 0.7 & 0.58 $\pm$ 0.02   & 96.7 $\pm$ 3.7 & 2.86 $\pm$ 0.08  \\ 
15~mJ cm$^{-2}$ PO (day 5) & mixed       & 74.16 $\pm$ 0.03 & 0.671 $\pm$ 0.002  & 0.23 $\pm$ 0.05        & 24.1 $\pm$ 0.2 & 0.61 $\pm$ 0.02   & 39.6 $\pm$ 1.5 & 1.04 $\pm$ 0.05  \\ 
10~mJ cm$^{-2}$ FO (day 5) & anisotropic & 72.3 $\pm$ 0.01  & 0.646 $\pm$ 0.001  & 0                      & 37.6 $\pm$ 0.4 & 0.813 $\pm$ 0.013 & 46.2 $\pm$ 0.9 & 1.65 $\pm$ 0.02  \\ \bottomrule[1pt]
\end{tabular}
\end{adjustbox}
\caption{\textbf{Extracted parameters from the anisotropic component of the 2D fit} Measurements are listed by pump fluence and scattering pattern. We distinguish between full overlap (FO) and partial overlap (PO) between probe and probe.  Extracted domain size $\pi/q$  (nm) and linewidth $\Gamma$ (nm$^{-1}$) from the pre-pumped signal, maximum shift in radial peak position (\%), maximum demagnetization (\%), quench time (ps) and quench speed (ps$^{-1}$) and demagnetization recovery time (ps) for all measurements.}
\label{table:fitted_params_aniso}
\end{table}

\section{Heating in samples due to high repetition rate}
\label{SI:heating}

The EuXFEL can achieve MHz repetition rates with the SCS beamline being able to deliver up to 150 X-ray pulses per train. However we observed that higher repetition rate and shorter pulse-to-pulse separation lead to visible damage damage in our samples. 

\begin{figure}[h]
\begin{center}
\includegraphics[width=\textwidth]{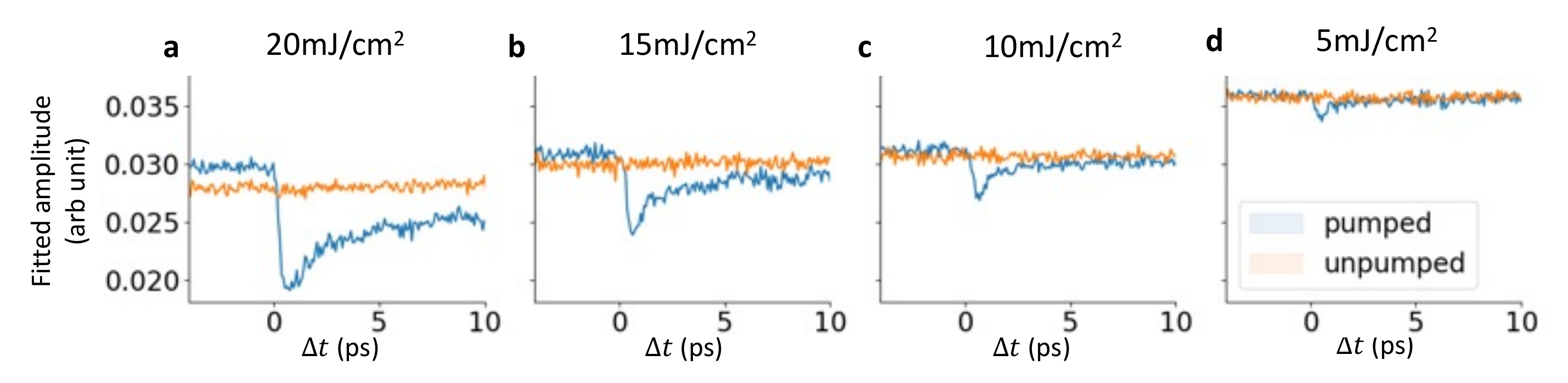}
\caption{\textbf{Slow heating effect of repetition rate} Fitted lobe amplitude for a series of pump-fluences \textbf{a} 20~mJ cm$^{-2}$, \textbf{b} 15~mJ cm$^{-2}$, \textbf{c} 10~mJ cm$^{-2}$, \textbf{d} 5~mJ cm$^{-2}$. We observe an offset between the pumped (blue) and unpumped (orange) data before $\Delta$t $\leq$ 0 attributed to the data not having fully returned to equilibrium between pulses. }
\label{fig:SI_heating}
\end{center}
\end{figure}

Our pumped-probe measurements were performed with 26 X-ray pulses per train at a repetition rate of 56~kHz and 13 IR pulses at half the repetition rate. Within one measurement we record both a pumped and an unpumped signal separated by 18~$\mu$s. While we observe no apparent damage in the unpumped signal we do observe an offset before time-zero in the fitted amplitudes of pumped and unpumped signals. \Cref{fig:SI_heating} shows that this offset increases with pump fluence. We conclude that 18~$\mu$s were not long enough for our samples to fully thermalize and return to equilibrium. We therefore have to keep in mind that our samples were at an elevated temperature and which could explain why the effects we are seeing are less pronounced than other measurements performed at different free electron facilities that run at considerably lower repetition rate (120~Hz at FERMI and LCLS).

\end{document}